\documentclass[prd,showpacs,letterpaper,twocolumn,nofootinbib,longbibliography,floatfix,superscriptaddress]{revtex4-2}

\usepackage{CJKutf8}

\usepackage{tikz}
\usetikzlibrary{arrows.meta, calc}
\usetikzlibrary{angles, quotes}
\usepackage{adjustbox}
\usepackage{amsmath}
\usepackage{cancel}
\usepackage{graphicx}
\usepackage{fullpage}
\usepackage[colorlinks=true,allcolors=blue]{hyperref}
\usepackage[capitalise]{cleveref}
\usepackage[utf8]{inputenc}
\usepackage{siunitx}
\usepackage{array}

\DeclareSIUnit \s {\second}
\DeclareSIUnit \ns {\nano\second}
\DeclareSIUnit \mus {\micro\second}
\DeclareSIUnit \ms {\milli\second}
\DeclareSIUnit \MB {\mega\byte}
\DeclareSIUnit \GB {\giga\byte}
\DeclareSIUnit \TB {\tera\byte}
\DeclareSIUnit \PB {\peta\byte}
\DeclareSIUnit \Mbps {\mega\bit/\s}
\DeclareSIUnit \Gbps {\giga\bit/\s}
\DeclareSIUnit \Tbps {\tera\bit/\s}
\DeclareSIUnit \Pbps {\peta\bit/\s}
\DeclareSIUnit \kton {\kilo\tonne} 
\DeclareSIUnit \kt {\kilo\tonne}
\DeclareSIUnit \Mt {\mega\tonne}
\DeclareSIUnit \eV {\electronvolt}
\DeclareSIUnit \keV {\kilo\electronvolt}
\DeclareSIUnit \MeV {\mega\electronvolt}
\DeclareSIUnit \GeV {\giga\electronvolt}
\DeclareSIUnit \TeV {\tera\electronvolt}
\DeclareSIUnit \PeV {\peta\electronvolt}
\DeclareSIUnit \EeV {\exa\electronvolt}
\DeclareSIUnit \m {\meter}
\DeclareSIUnit \cm {\centi\meter}
\DeclareSIUnit \in {\inchcommand}
\DeclareSIUnit \km {\kilo\meter}
\DeclareSIUnit \kV {\kilo\volt}
\DeclareSIUnit \kW {\kilo\watt}
\DeclareSIUnit \MW {\mega\watt}
\DeclareSIUnit \MHz {\mega\hertz}
\DeclareSIUnit \mrad {\milli\radian}
\DeclareSIUnit \year {years}
\DeclareSIUnit \POT {POT}
\DeclareSIUnit \sig {$\sigma$}
\DeclareSIUnit\parsec{pc}
\DeclareSIUnit\lightyear{ly}
\DeclareSIUnit\foot{ft}
\DeclareSIUnit\ft{ft}
\DeclareSIUnit \ppb{ppb}
\DeclareSIUnit \ppt{ppt}
\DeclareSIUnit \samples{S}
\DeclareSIUnit \pe{PE}

\usepackage{pifont}
\usepackage{enumitem}
\usepackage{url}
\usepackage{rotating}
\usepackage{epsfig,graphics,rotate}
\usepackage{flushend}
\usepackage{bm}
\usepackage{amssymb}
\usepackage{amsmath}
\usepackage{amsfonts}
\usepackage{gensymb}
\usepackage{booktabs}
\usepackage{longtable}   
\usepackage{colortbl}    

\usepackage{microtype}
\usepackage{multirow, makecell}
\usepackage{lipsum}
\usepackage{fullpage}
\usepackage[normalem]{ulem}

\usepackage{xspace}
\usepackage{comment}
\usepackage{floatrow}
\usepackage{ragged2e}

\usepackage{nicefrac, xfrac}
\usepackage{mathtools} 
\usepackage{relsize}   

\usepackage{float}

\definecolor{claude}{RGB}{214,98,0}

\renewcommand{\arraystretch}{1.15}

\definecolor{lime}{HTML}{A6CE39}
\DeclareRobustCommand{\orcidicon}{
	\begin{tikzpicture}
	\draw[lime, fill=lime] (0,0)
	circle [radius=0.16]
	node[white] {{\fontfamily{qag}\selectfont \tiny ID}};
	\draw[white, fill=white] (-0.0665,0.095)
	circle [radius=0.005];
	\end{tikzpicture}
	\hspace{-2mm}
}

\foreach \x in {A, ..., Z}{\expandafter\xdef\csname orcid\x\endcsname{\noexpand\href{https://orcid.org/\csname orcidauthor\x\endcsname}
			{\noexpand\orcidicon}}
   }

\newcommand{\dCP}{\delta_{\rm{CP}}}
\newcommand{\dmsq}{\Delta m^2_{31}}
\newcommand{\ssqtt}{\sin^2\theta_{23}}

\newcommand{\DeepCore}{IceCube-DeepCore\xspace}

\newcommand{\SK}{Super-Kamiokande\xspace}
\newcommand{\etab}{\vec \eta}
\newcommand{\thetab}{\vec \theta}

\newcommand{\ktune}{\mathcal{K}}
\newcommand{\NEUT}{\mathrm{NEUT}}
\newcommand{\GENIE}{\mathrm{GENIE}}

\begin{document}

\begin{CJK*}{UTF8}{gbsn}


\title{Evidence for Normal Neutrino Ordering \\ 
with Combined Atmospheric Neutrino Experiments}

\author{Miaochen Jin (靳淼辰)\orcidD{}}
\email{miaochenjin@g.harvard.edu}
\affiliation{Department of Physics \& Laboratory for Particle Physics and Cosmology, Harvard University, Cambridge, MA 02138, USA}

\author{Carlos A.~Arg{\"u}elles\orcidA{}}
\email{carguelles@g.harvard.edu}
\affiliation{Department of Physics \& Laboratory for Particle Physics and Cosmology, Harvard University, Cambridge, MA 02138, USA}

\author{\\ Pablo Fern\'andez-Men\'endez\orcidB{}}
\email{pablo.fer@cern.ch}
\affiliation{Universidade de Santiago de Compostela, Instituto Galego de F\'isica de Altas Enerx\'ias (IGFAE), R\'ua de Xoaqu\'in D\'iaz de R\'abago, s/n, Santiago de Compostela, 15705, Spain}

\author{Iv\'an Mart\'inez-Soler\orcidC{}}
\email{ivan.martinez@uam.es}
\affiliation{Departamento de F\'isica Te\'orica, Universidad Aut\'onoma de Madrid, Cantoblanco, 28049, Madrid, Spain}
\affiliation{Institute for Particle Physics Phenomenology, Department of Physics, Durham University, Durham DH1 3LE, UK}

\date{\today}

\begin{abstract}
We present a combined analysis of the latest publicly available atmospheric neutrino data from Super-Kamiokande
including all data-taking stages previous to the gadolinium phase, IceCube-DeepCore, and the six-detection-unit configuration of KM3NeT/ORCA.
Our fit uses all event samples simultaneously and treats the systematic uncertainties shared among experiments as fully correlated, while detector-response uncertainties are modeled independently for each experiment.
The combined data constrain the atmospheric oscillation parameters to $\Delta m^2_{31} = 2.44^{+0.06}_{-0.04} \times 10^{-3}\,\mathrm{eV}^2$ and $\sin^2\theta_{23} = 0.52^{+0.03}_{-0.04}$, with a precision of $2.2\%$ and $6.1\%$, respectively, which is competitive with current accelerator-based measurements, and independently constrain the $CP$-violating phase to $\delta_{CP} = 1.30^{+0.24}_{-0.28}~\pi$.
Finally, comparing the orderings of the neutrino mass spectrum, we find a preference for the normal ordering with $\Delta\chi^2 = 11.0$; further calibrated with pseudoexperiments, we exclude the inverted ordering at $99.95\%$ confidence level ($3.3\sigma$).
In a joint fit including 207 days of JUNO data, we further exclude the inverted ordering at $\Delta \chi^2 = 15.1$, corresponding to $3.9 \sigma$.
\end{abstract}
\maketitle
\end{CJK*}


\section{Introduction}\label{sec:intro}
Atmospheric neutrinos have played a central role in establishing that neutrinos have mass and that their flavor and mass states are misaligned~\cite{Lipari:2018qqz,Koshio:2020yex}.
First hints of a deficit of muon neutrinos in IMB~\cite{PhysRevLett.69.1010} and Kamiokande~\cite{Kamiokande-II:1992hns} were confirmed by Super-Kamiokande's discovery of atmospheric neutrino oscillations~\cite{Super-Kamiokande:1998kpq}, an observation now understood as the flavor evolution of massive neutrinos~\cite{deGouvea:2016qpx}.
This evolution depends on the ratio of the source-to-detector distance, $L$, to the neutrino energy, $E_\nu$, and atmospheric neutrinos cover this ratio over eight orders of magnitude.
This unmatched coverage of $L/E_\nu$, together with trajectories that traverse the largest matter densities accessible to any neutrino experiment~\cite{Wolfenstein:1977ue}, makes atmospheric neutrinos precision instruments for oscillation physics~\cite{Esteban:2018azc,deSalas:2020pgw,Arguelles:2022hrt,Olavarrieta:2024eaq,Capozzi:2025wyn} and powerful probes of new physics as reported in~\cite{IceCube:2020tka,IceCube:2017qyp,Gonzalez-Garcia:2004pka,KM3NeT:2024fkk,SuperKamiokande:2011dam,SuperKamiokande:2014gyp,Sponsler:2024iej} to name a few.

This wide range in $L/E_{\nu}$, combined with Earth matter effects, gives atmospheric neutrinos a remarkable sensitivity to all the remaining unknown oscillation parameters of the three-flavor paradigm.
The mixing angle between second and third neutrino generations, $\theta_{23}$, is the least precisely known angle and its octant remains unresolved; the ordering of the neutrino mass spectrum --- whether $\nu_3$ is heavier (normal ordering, NO) or lighter (inverted ordering, IO) than the other two states --- is undetermined; and the $CP$-violating phase, $\dCP$, is only weakly constrained. Current long-baseline measurements by T2K~\cite{T2K:2021xwb} and NOvA~\cite{NOvA:2019cyt} are in tension~\cite{T2K:2025wet}, which limits the ability of present data to reduce these uncertainties and carries the weight to the next-generation of accelerator experiments DUNE~\cite{PhysRevD.105.072006} and Hyper-Kamiokande~\cite{Hyper-Kamiokande_Collaboration2026-ff}.

Of these questions, the determination of the neutrino mass ordering has important implications for uncovering the potential Majorana nature of the neutrino~\cite{Agostini:2022zub,KamLAND-Zen:2024eml,Furry:1939qr,LEGEND:2025jwu}.
Additionally, its determination has direct consequences in cosmological measurements~\cite{Lesgourgues:2006nd,Elbers:2025vlz,Jiang:2024viw,Naredo-Tuero:2024sgf}, which currently prefer normal ordering.
Moreover, a non-zero $CP$-violating phase provides insight into the Universe's matter-antimatter  asymmetry~\cite{Sakharov:1967dj,Kuzmin:1985mm,Fukugita:1986hr}.
This article answers the former question and provides new constraints on the latter.

Atmospheric neutrino experiments offer an independent handle on all these parameters simultaneously.
A full combined analysis of atmospheric neutrino experiments properly accounting for the correlation of their shared systematic uncertainties, yields a comprehensive picture of current atmospheric neutrino measurements, significantly reducing the remaining uncertainties on the three-flavor oscillation parameters~\cite{Arguelles:2022hrt}. 



In this article, we present a combined analysis of all the publicly available atmospheric neutrino data in the standard three-neutrino framework, including \SK in all of its data-taking stages~\cite{Super-Kamiokande:2023ahc} with ultra-pure water, the latest \DeepCore data release~\cite{IceCubeCollaboration:2023wtb}, and the data collected by KM3NeT/ORCA in its six-detection-unit configuration (ORCA6)~\cite{km3netcollaboration2024measurementneutrinooscillationparameters}.
These three experiments observe the same flux through the same interactions, and our fit exploits this by treating flux and cross-section uncertainties as fully correlated across experiments, while detector-response uncertainties remain experiment-specific; the systematic treatment is analogous to the one introduced in Refs.~\cite{Arguelles:2022hrt,Jin:2026cpt} and is described in~\cref{sec:analysis}.
From this combination we obtain measurements of $\dmsq$ and $\ssqtt$ that are competitive with accelerator-based results, constraints on $\dCP$ that are independent of beam experiments, and a strong preference for the normal ordering as we will quantify in~\cref{sec:results}.

\begin{figure}
    \centering
    \includegraphics[width=\linewidth]{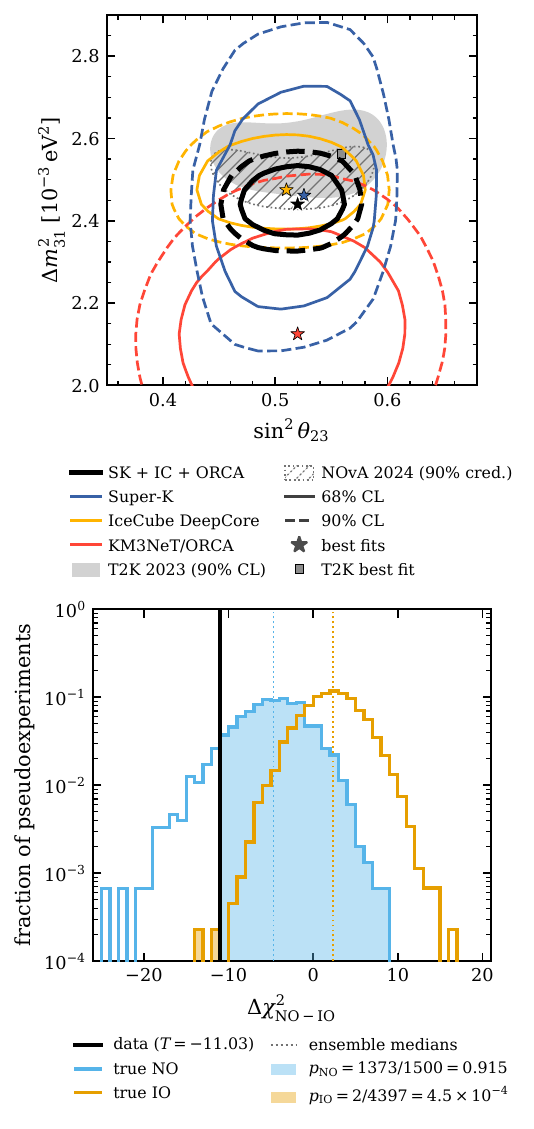}
    \caption{\textbf{\textit{Major Results on oscillation parameters and neutrino mass ordering.}} \textbf{Top panel:} Individual experiments' $68\%$ and $90\%$ allowed region contours in $(\Delta m^2_{31}, \, \sin^2 \theta_{23})$ and the combined fit. All results reproduced by this work. Overlaid with $90\%$ contours from officially released accelerator results~\cite{T2K:2023smv,T2K:2023data,NOvA:2024data,NOvA:2025tmb}.
    For solar parameters, we have assumed $\Delta m^2_{21}=7.54\times10^{-5}$\,eV$^2$, $\sin^2\theta_{12} = 0.3088$ and $\sin^2\theta_{13} = 0.022$.
    \textbf{Bottom Panel}: Distribution of $T = \chi^2_{\rm NO} - \chi^2_{\rm IO}$ (negative values favor the normal ordering) for ensembles of pseudoexperiments generated under a true normal ordering (blue) and a true inverted ordering (orange).}
    \label{fig:headline}
\end{figure}

The combination of the three experiments yields a precise determination of the atmospheric mass splitting, $\Delta m^2_{31} = 2.44\times 10^{-3}\,\text{eV}^2$, with a relative precision of $2.2\%$, comparable to the $\sim 1\%$ precision reached by long-baseline experiments~\cite{Esteban:2018azc}.
We find the mixing angle to be close to maximal mixing within the $1\sigma$ region of the best-fit value of $\sin^2\theta_{23} = 0.52$, and with no strong preference for the higher octant, with only $\Delta \chi^2 = 0.3$ preference over the mirror parameter in the lower octant.
The best-fit value of the CP-violating phase is $\dCP=1.30\pi^{+0.24\pi}_{-0.28\pi}$, consistent with combined long-baseline results~\cite{T2K:2025wet} and further supporting their disfavoring of small $\dCP$ values.
Regarding the mass ordering, we find a preference for NO with $\Delta\chi^2 = 11.0$, and in this case where Wilks' theorem does not apply, it is calibrated to $3.3 \sigma$ using pseudoexperiments. These major results are shown in \cref{fig:headline} and discussed further in \cref{sec:results}. 

Finally, we combine our atmospheric-neutrino analysis with the recent JUNO results based on 207 days of data taking. This joint analysis further strengthens the exclusion of the inverted mass ordering, reaching a significance of $3.9\sigma$. The combined fit is discussed in \cref{sec:JUNO} and \cref{app:juno}.

Our work is a significant leap forward compared to a recent work~\cite{eller2026atmosphericneutrinooscillationspicture} that also analyzes the atmospheric data. This is because our work uses all the data release information simultaneously with 930 bins, an independently developed high-statistics SK MC set, a bug-fixed version of the SK data release, extensive statistical studies of the neutrino mass ordering preference, as well as a joint fit with recent JUNO results~\cite{JUNO:2026tap}.

The rest of this article is organized as follows.
In~\cref{sec:data}, we describe the three experiments and the public data sets used in this analysis.
In~\cref{sec:analysis}, we present the analysis methodology, the correlated treatment of the flux and cross-section uncertainties, the detector-response uncertainties, and the statistical methods.
In~\cref{sec:results}, we report the measured oscillation parameters, the octant and ordering preferences, and compare them with accelerator results. In~\cref{sec:JUNO} we discuss the joint fit with the JUNO reactor results.
Finally, we conclude in~\cref{sec:conclusions}.

\section{Atmospheric Neutrino Simulation and Data Sets}\label{sec:data}
Our combination rests on the three water(ice)-Cherenkov detectors that measure the atmospheric neutrino flux at overlapping energies with complementary strengths.
\SK provides finely classified event samples from 100\,\si{\MeV} to energies up to 100~\si{\GeV}, including statistical neutrino-antineutrino separation based on hydrogen neutron-tagging; \DeepCore provides the largest sample of events in the 5--100\,\si{\GeV} range, where the first oscillation minimum lies; and ORCA6, although statistically smaller and still under construction, adds an independent measurement covering the same energy range as \DeepCore and extending down to 2\,\si{\GeV}, using data collected by the six optical arrays already deployed.
The three experiments and the public data sets used in this work are described below

\subsection{Super-Kamiokande}\label{sec:SK}

Super-Kamiokande is a 50\,\si{\kton} water-Cherenkov detector located in the Kamioka mine, Japan, instrumented with more than 11{,}000 inward-facing 20-inch PMTs~\cite{FUKUDA2003418}.
We use the atmospheric neutrino data collected across all data-taking stages, SK-I through SK-V, corresponding to 6511.3 live days and an exposure of 484.2\,\si{\kton}-years with the expanded fiducial volume~\cite{Super-Kamiokande:2023ahc}.
Following the official analysis, events are classified as fully contained (FC), partially contained (PC), and upward-going muons (Up-$\mu$), with FC events further divided by their number of rings, reconstructed visible energy (Sub-GeV or Multi-GeV), and particle identification ($e$-like or $\mu$-like).
In SK-IV and SK-V, improved electronics enable neutron tagging via the detection of the 2.2\,\si{\MeV} photon produced from neutron capture on hydrogen, and the corresponding samples are further divided by the number of tagged neutrons, improving the statistical separation between neutrinos and antineutrinos~\cite{FernandezMenendez:2021jfk,Super-Kamiokande:2023ahc}.

The Monte Carlo simulation used in this work inherits from that developed in Ref.~\cite{Arguelles:2022hrt}, with GENIE~\cite{Andreopoulos:2009rq} as the interaction generator and adds in corrections to match the distributions of the simulation provided in the collaboration's release~\cite{super_kamiokande_collaboration_2023_8401262}. To obtain these, the statistical quantities of the logarithm of the true energy and the cosine of the zenith angle distributions for each sample, bin, and interaction group are fit to a piecewise cubic polynomial that closely matches their cumulative distributions.
These corrections allow us to improve the reproduction of sample selection efficiencies and of energy and direction reconstruction of the detector.

Additionally, in the process of developing this work, we noticed a bug in the SK data release~\cite{super_kamiokande_collaboration_2023_8401262} where the true energy quantiles show ramped-up behaviors, localized later as binning artifacts for the high-energy Partially-Contained (PC) and up-going muon (Up-$\mu$) samples. This is confirmed by the SK Collaboration through private correspondence,\footnote{Email communications with the Super-Kamiokande Collaboration, August 11th, 2026} which provided corrected true-energy quantiles that will be released publicly; we are using a preview version for this article.

\subsection{IceCube-DeepCore}\label{sec:IC}

IceCube is a cubic-kilometer ice-Cherenkov detector at the South Pole, consisting of 5160 digital optical modules deployed at depths between 1450\,\si{\m} and 2450\,\si{\m}~\cite{IceCube:2006tjp}.
Its denser inner sub-array, DeepCore, lowers the energy threshold to around 5\,\si{\GeV}, giving access to the energies where the muon-disappearance minimum for trans-Earth baselines occurs~\cite{IceCube:2019dqi}.
%
We use the most recent public data release~\cite{IceCubeCollaboration:2023wtb}.
It corresponds to 8 years of DeepCore data and contains up-going events distributed in logarithmic bins of reconstructed energy between 6.31 and 158.49\,\si{\GeV}.
Events are classified into tracks, dominated by $\nu_\mu$ charged-current (CC) interactions, and mixed events, which consist mostly of $\nu_\mu$ CC interactions with an approximately $30\%$ contamination from other flavors.
The release provides per-bin Monte Carlo expectations and detector-response variations, which we incorporate as described in~\cref{sec:analysis}.

\subsection{KM3NeT/ORCA6}\label{sec:ORCA}

KM3NeT/ORCA is a megaton-scale water-Cherenkov detector under construction in the Mediterranean Sea, optimized for atmospheric neutrinos in the 1--100\,\si{\GeV} range~\cite{KM3Net:2016zxf}.
We use the data collected with its first six detection units (ORCA6), corresponding to 510 days of livetime and an exposure of 433\,\si{\kton}-years~\cite{km3netcollaboration2024measurementneutrinooscillationparameters}.
The ORCA6 sample contains 5828 events after selection, classified into high-purity tracks, low-purity tracks, and showers, and binned in reconstructed energy and cosine of the zenith angle.
The public release ships per-event Monte Carlo weights and statistical variances, which are used in generating pseudoexperiments for assessing goodness-of-fit, see~\cref{sec:analysis} and~\cref{app:fit}.

\begin{figure*}[t]
\centering
\includegraphics[width=\textwidth]{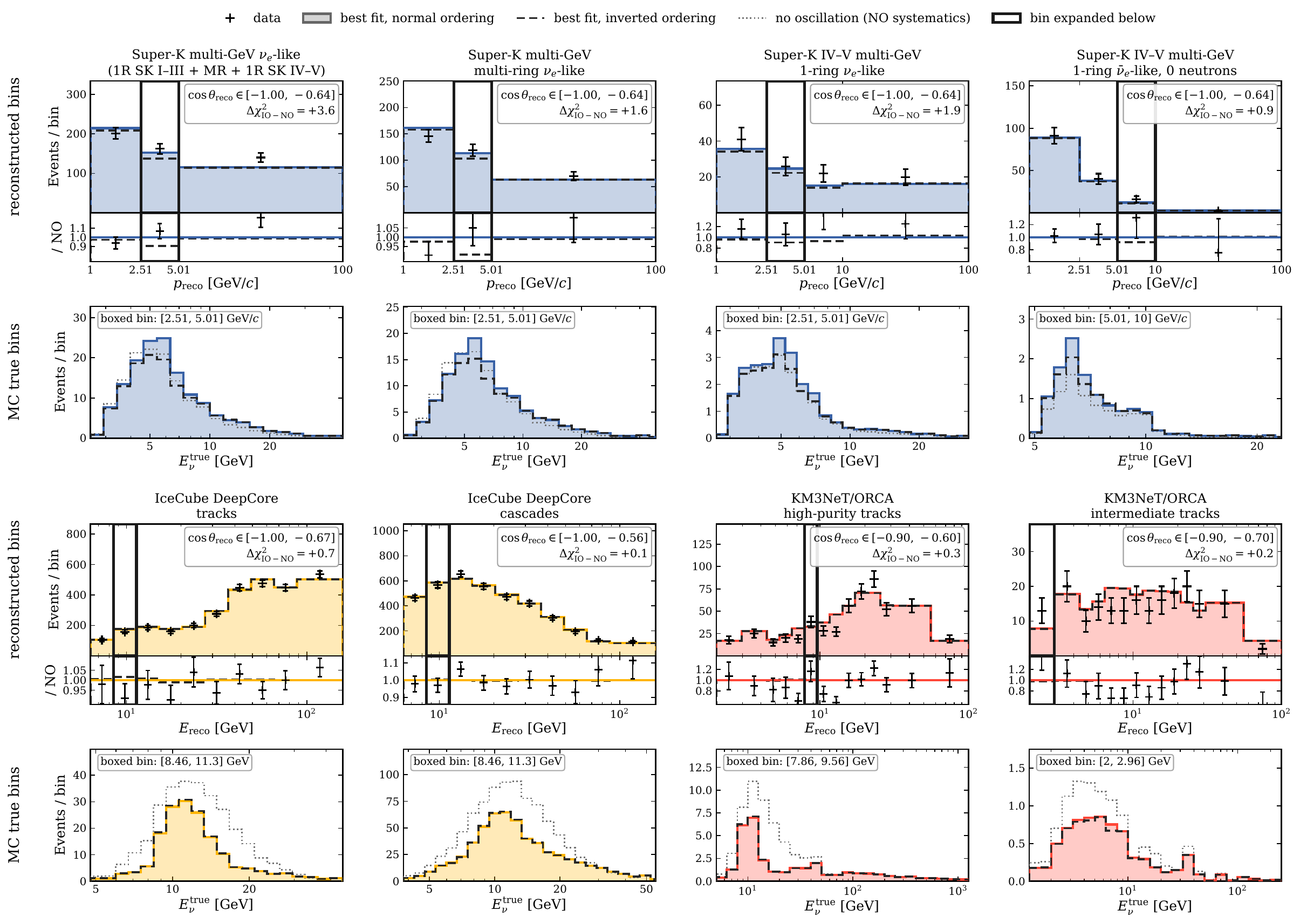}
\caption{\textbf{\textit{Observed event distributions compared with the combined-fit prediction at NO and IO best fit points.}}
We show the 4 top NMO sensitivity contributor sample from SK and 2 each from ORCA6 and IceCube-DeepCore, focusing on the core-crossing cosine zenith bins. In each sample's top-down pair, the upper panel shows the reconstructed variable space distributions, where a boxed reconstructed bin is expanded in the lower panel in true variable distribution. The colored histograms are the distributions at best fit NO parameters; the dashed histograms are the distributions at best fit IO parameters, and the dotted are at no oscillations as reference.
\label{fig:eventdist}}
\end{figure*}

\section{Analysis Methods and Systematic Uncertainties}\label{sec:analysis}
Our analysis fits all event samples of the three experiments simultaneously, sharing the nuisance parameters that describe physics common to all of them.
The atmospheric neutrino flux and the neutrino-water cross sections are identical for \SK, \DeepCore, and ORCA6, so the associated uncertainties are treated as fully correlated across experiments; only the detector-response uncertainties are experiment-specific.
This treatment allows the high-statistics samples of one experiment to constrain nuisance parameters that would otherwise degrade the sensitivity of another. This is the key methodological difference from previous analyses of atmospheric neutrino data, in which only the published likelihoods were combined.
This approach is especially relevant in the 1 to 10~GeV region where all three experiments share a significant fraction of their statistics and data is most sensitive to the neutrino mass ordering, the $\theta_{23}$ mixing angle and, to a lesser extent, the CP phase.
We use the \texttt{Pynu} framework~\cite{Arguelles:2022hrt,Olavarrieta:2024eaq,Jin:2026cpt} throughout.

\subsection{Oscillation calculation and parameters}\label{sec:osccalc}

We compute the neutrino evolution through the Earth numerically with \texttt{nuSQuIDS}~\cite{Arguelles:2021twb}, using the Preliminary Reference Earth Model~\cite{Dziewonski:1981xy} for the matter and electron densities and averaging the mixing probabilities in those regimes where fast oscillations occur.
The solar parameters are fixed to $\Delta m^2_{21} = 7.54\times 10^{-5}\,\mathrm{eV}^2$ and $\sin^2\theta_{12} = 0.3088$~\cite{Esteban:2026phq}, to which atmospheric data have negligible sensitivity. Provided the precise measurement from reactor experiments, we focus on the scenario with $\sin^2\theta_{13}$ fixed to the reactor value~\cite{DayaBay:2018yms}.

\subsection{Atmospheric flux uncertainties}\label{sec:fluxsys}

The atmospheric neutrino flux prediction is common to the three experiments, and so are its uncertainties.
We take the HKKM2014 calculation~\cite{Honda:2015fha}, interpolated with \texttt{NuFlux}~\cite{nuflux}, as the central flux model, and parametrize its uncertainties following Refs.~\cite{Barr:2006it, Abe_2018, Fedynitch:2022vty}, with special attention below a few-GeV where the flux uncertainties are largest and most relevant for the measurement of the $CP$-violating phase.
These ranges assume budgets from recent \SK~\cite{Super-Kamiokande:2023ahc} and IceCube~\cite{IceCube:2024xjj} analyses and are summarized in~\cref{table:FluxSysts}.

\subsection{Cross-section uncertainties}\label{sec:xsecsys}

Neutrino-water interaction uncertainties are likewise shared by all three detectors.
The relevant channels span charged-current quasi-elastic scattering (CCQE 1p1h and 2p2h), which dominates the \SK Sub-GeV samples where the $\dCP$ sensitivity resides; CC resonant meson production (mostly CC1$\pi$), relevant for all experiments between $\sim$1 and 4\,\si{\GeV}; and CC deep-inelastic scattering (CC DIS), which dominates the \DeepCore and ORCA6 samples. These last two interaction modes are particularly important for this analysis because they cover the energy region with the most overlap among experiments and carry the sensitivity to $\dmsq$ and the mass ordering. Due to its larger uncertainty, CC$\nu_\tau$ cross section is parametrized separately. 
Additionally, uncertainties coming from neutral-current (NC) interactions are included with a coarser parametrization.
The 1$\sigma$ priors, listed in~\cref{table:XSecSysts}, follow the official oscillation analyses of the three experiments~\cite{Super-Kamiokande:2023ahc,IceCubeCollaboration:2023wtb,km3netcollaboration2024measurementneutrinooscillationparameters}. 

\subsection{Detector uncertainties}\label{sec:detsys}

Detector-response uncertainties are uncorrelated among experiments and follow the public information released by each collaboration.
For \SK, we implement as closely as allowed by our simulation the detector systematics described in the official analyses~\cite{Super-Kamiokande:2023ahc, Abe_2018} separately for the SK-I, SK-II, SK-III and SK-IV to SK-V periods. 
For \DeepCore, we use the detector-response parametrization shipped with the data release~\cite{IceCubeCollaboration:2023wtb}, which includes the optical efficiency of the modules, the bulk-ice absorption and scattering, the hole-ice angular acceptance, and the atmospheric-muon background normalization; these are implemented as hyperplane interpolations between discrete detector Monte Carlo sets.
For ORCA6, we follow the treatment of the collaboration's analysis~\cite{km3netcollaboration2024measurementneutrinooscillationparameters}, with free normalizations for each event class and the muon-background normalization. The KM3NeT Collaboration does not release a treatment for the energy scale as a detector systematics, so we leave it out of the fit, where the effect on the combined fit is small and discussed further in \cref{app:fit}.
A complete list of nuisance parameters, priors, and pulls at the best-fit point is provided in the Supplemental Material (\Cref{app:systs}).

\subsection{Statistical methods}\label{sec:stats}

We compare data and prediction with a binned Poisson likelihood.
The \SK prediction is computed from our updated high-statistics Monte Carlo simulation with 100 times the data exposure, resulting in negligible statistical uncertainty across all analysis bins. By contrast, the effect of the finite Monte Carlo statistics of the \DeepCore and ORCA6 releases on the fitted parameters is noticeable but small: replacing the bare Poisson term by the Barlow--Beeston ``lite'' likelihood~\cite{BARLOW1993219,Conway:2011in} moves the best-fit values of the oscillation parameters by less than $0.2\sigma$ and changes the widths of their $1\sigma$ intervals by less than $5\%$. Accounting for the finite MC statistics does have an effect on the goodness-of-fit, which will be discussed later in \cref{sec:conclusions} and \cref{app:fit}. Dropping the per-bin scale factors also keeps the analytic gradient of the test statistic with respect to every nuisance parameter exact.

Writing $O_i$ for the observed count and $E_i(\bm\eta)$ for the total expectation given nuisance parameters $\bm\eta$,  the test statistic minimized at each point of the oscillation-parameter grid reads

\begin{widetext}
\begin{equation}
\label{eq:chi2-poisson}
\chi^2(\thetab,\etab)
  = 2\!\sum_{{\rm Exp}}\sum_{i\in{\rm Bins}}\!
        \left[\,E_i(\thetab,\etab) - O_i
              + O_i\,\ln\!\frac{O_i}{E_i(\thetab,\etab)}\right]
  + \sum_{j\in{\rm Syst}}\!\left(\frac{\tilde\eta_j-\eta_j}{\sigma_j}\right)^{\!2}.
\end{equation}
\end{widetext}

The data fit is performed with the software \texttt{PyNu}~\cite{pynu}, which also computes analytically the gradient of the $\chi^2$ improving the robustness and performance of the fit.

In performing the data fit, we adopt a binning approach that avoids the event-by-event reweighting, thereby accelerating the fit process. This logic has been employed by IceCube~\cite{IceCube:2025gollumfit}. In presence of high-statistics MC sets in our fit, such a treatment accelerates minimization by factors of $30\times$ to $300\times$ depending on the minimizer. We use the binning and response with SK as the example experiment as it has the largest number of binning axes. Based on the event-by-event Monte Carlo, we build a bare response kernel at nominal nuisance and no oscillations $R[i |e,k,E,Z]$, where $i$ labels the analysis reconstructed bin number (930 bins in total), $e$ labels the Super-Kamiokande detector phase, $k$ labels the interaction mode, and $(E, Z)$ labels the true energy and zenith bins. Aside from this bare kernel, in order to faithfully match the simulated MC rates to the officially released per-bin rates and adjust for small deviations from modeling, we can define tune kernels to adjust the bare kernel at nominal systematics and no oscillation. We discuss this in detail in \cref{app:MC}.

We then construct the response tensors at oscillation parameters and nuisance parameters, independent of the event MC from oscillation software and analytic definitions. This includes the oscillation response $\phi_f(E,Z,\vec \theta)$ where $f$ is per flavor and $\vec \theta$ is the physics parameters, the true-space nuisance response tensor $w_k(E; \vec \eta)$ for nuisance parameters $\vec \eta$, and the reco-space nuisance parameters $\mathcal{T}_e$ for energy migration and $D_{e, i}$ for detector-era-related systematics. The forward computation of expected bin count is given by

\begin{widetext}
    \begin{equation}
E_i(\thetab,\etab)\;=\;\sum_e D_{e,i}(\etab)\,
\big[\mathcal T_e(\etab)\big]_i\Big(\sum_{k,E,Z}R[i|e,k,E,Z]\;\phi_k(E,Z;\vec \theta)\;w_k(E;\etab)\Big),
\label{eq:chain}
\end{equation}
\end{widetext}

Confidence regions are constructed assuming Wilks' theorem~\cite{Algeri:2019lah} for the oscillation parameters; the coverage of the mass-ordering test statistic, which compares non-nested hypotheses, uses pseudoexperiments instead. The pseudoexperiment datasets are generated by fluctuating the nominal expectation according to the statistical and prior systematic uncertainties and is further discussed in~\cref{sec:results}.

\section{Results}\label{sec:results}

\begin{table*}
    \centering
\begin{tabular}{c||c|c|c|c||c|c|c|c}
Parameter & SuperK & IceCube & ORCA6 & \textbf{Combined} & SuperK & IceCube & ORCA6 & \textbf{Combined} \\\hline\hline
$\dmsq [10^{-3}\,\mathrm{eV}^2]$ & $2.47^{+0.13}_{-0.19}$ & $2.48^{+0.07}_{-0.04}$ & $2.13^{+0.16}_{-0.17}$ & $2.44^{+0.06}_{-0.04}$ & $-2.26^{+0.21}_{-0.17}$ & $-2.40^{+0.07}_{-0.04}$ & $-2.02^{+0.17}_{-0.17}$ & $-2.33^{+0.05}_{-0.06}$\\
$\ssqtt$ & $0.53^{+0.05}_{-0.05}$ & $0.51^{+0.04}_{-0.05}$ & $0.52^{+0.06}_{-0.07}$ & $0.52^{+0.03}_{-0.04}$ & $0.53^{+0.04}_{-0.06}$ & $0.50^{+0.06}_{-0.04}$ & $0.52^{+0.06}_{-0.07}$ & $0.52^{+0.03}_{-0.04}$\\
$\dCP$ & $1.25\pi^{+0.28\pi}_{-0.28\pi}$ & $1.70\pi$\,(---) & $1.70\pi$\,(---) & $1.30\pi^{+0.24\pi}_{-0.28\pi}$ & $1.20\pi^{+0.28\pi}_{-0.30\pi}$ & $0.80\pi$\,(---) & $0.70\pi$\,(---) & $1.20\pi^{+0.26\pi}_{-0.31\pi}$ \\
\hline
$\chi^2/$ndof & $991.4/930$ & $207.7/200$ & $507.8/450$ & $1736.8/1580$ & $997.6/930$ & $209.4/200$ & $508.3/450$ & $1747.8/1580$\\ \hline
\multicolumn{1}{c}{} & \multicolumn{4}{c}{\emph{\textbf{Normal Ordering}}}  & \multicolumn{4}{c}{\emph{\textbf{Inverted Ordering}}} \\
\end{tabular}
    \caption{\textbf{\textit{Best-fit values and $1\sigma$ intervals from the individual and combined fits.}}
    Normal ordering (left) and inverted ordering (right), with $\sin^2\theta_{13}=0.022$ fixed. Intervals are $\Delta\chi^2=1$ crossings of the one-dimensional profiles, with all other oscillation and nuisance parameters profiled; $\dCP$ is in units of $\pi$. For inverted ordering $\dmsq<0$, and the upper (lower) uncertainty moves towards smaller (larger) $|\dmsq|$. The \DeepCore and ORCA6 profiles in $\dCP$ are flat, with $\Delta\chi^2<0.4$ over the full range: their best-fit values are quoted, but no $1\sigma$ interval exists (---). }
    \label{tab:results_13fix}
\end{table*}

\begin{figure*}[ht]
\centering
\includegraphics[width=\textwidth]{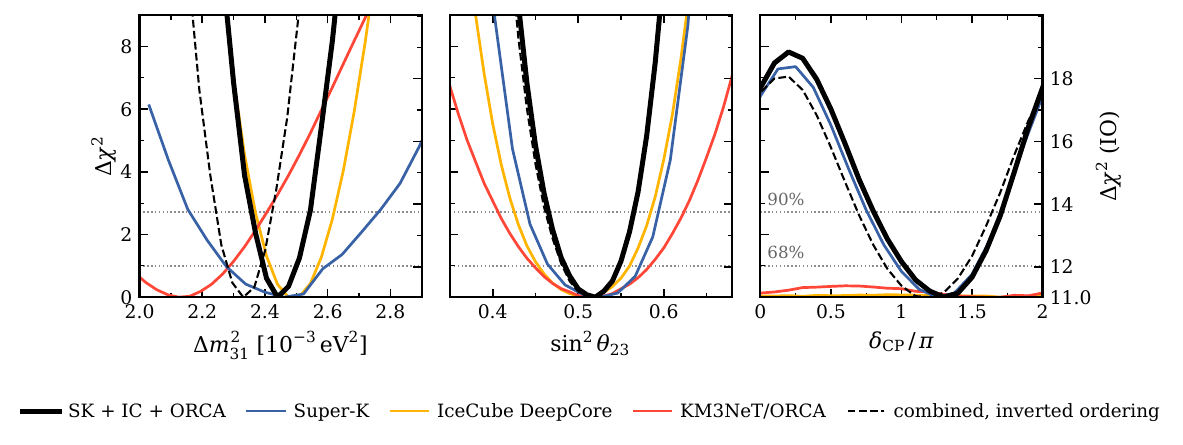}
\caption{\textbf{\textit{Normal-ordering constraints from the combined atmospheric-neutrino fit.}}
Results from the combined analysis of \SK, \DeepCore, and ORCA6 (thick black) compared with the corresponding individual fits of \SK (blue), \DeepCore (yellow), and ORCA6 (red) alone.
We show one-dimensional $\Delta\chi^2$ profiles in $\dmsq$, $\ssqtt$ and $\dCP$, each minimized over the other two parameters and taken relative to the minimum of the respective fit; dotted lines mark $\Delta\chi^2=1$ and $2.71$ (68\% and 90\% CL for one degree of freedom).
The combined fit gives $\dmsq = 2.44^{+0.06}_{-0.04}\times10^{-3}$\,eV$^2$, $\ssqtt = 0.52^{+0.03}_{-0.04}$ and $\dCP = 1.30\pi^{+0.24\pi}_{-0.28\pi}$.
\label{fig:results}. The allowed ranges for the corresponding parameters under the inverted-ordering hypothesis are shown by the dashed black lines. The right vertical axis shows the $\Delta\chi^2$ for the inverted-ordering scan.}
\end{figure*}

The combined fit yields a good description of the three experiments data within the three-flavor neutrino scenario. We note a remaining residual tension for ORCA6, found to be largely driven by the limited Monte Carlo statistics of the public data release. \Cref{fig:eventdist} compares the observed event distributions in several relevant samples for the analysis with the prediction at the combined best-fit point assuming both mass orderings.

The goodness of fit is $\chi^2/\mathrm{ndof} = 1736.8/1580$, with $p = 0.011$ as estimated from pseudoexperiments generated at the best fit and analyszd with the full fit. This low $p$-value is driven by ORCA6 which has $p = 0.011$ at the combined best fit. \SK and \DeepCore p-values are respectively 0.164 and 0.136 at the combined best fit.
Taking into account the finite MC statistics in addition to Poisson statistical fluctuations when throwing pseudoexperiment realizations, the combined as well as individual goodness of fit p-values increase; more discussions are provided in \cref{app:fit}.

We further assess the compatibility of the three data sets with the shared systematics model. The pulls of the 39 correlated flux and cross-section nuisance parameters are all within $2.5\sigma$ of their priors, the largest being that of the CCQE normalization ($-2.5\sigma$), with other correlated nuisance parameters pulling below $2\sigma$. The full table is provided in \cref{app:systs}.

Finally, we quantify the consistency among the three experiments with a parameter-goodness-of-fit test~\cite{Maltoni:2003cu}, which compares the minimum of the combined fit with the sum of the minima of the three individual fits and is calibrated with the same pseudoexperiments. We obtain $\chi^2_{\rm PG} = 29.1$ and $p = 0.014$ ($2.2\sigma$). 
Owing to the treatment of shared systematic uncertainties, the PG-test statistic does not necessarily follow the conventional $\chi^2$ distribution with degrees of freedom determined by simple parameter counting. Instead, we characterize its distribution using pseudoexperiments and find that it is approximately described by a $\chi^2$ distribution with 15 effective degrees of freedom. This effective number of degrees of freedom accounts for the impact of the shared systematic uncertainties on the distribution of the test statistic.
The residual tension is carried by \DeepCore and ORCA6, whose data are described worse by $\Delta\chi^2 = 13.7$ and $9.6$ at the combined best fit than at their own, while \SK changes by only $2.3$.


\Cref{tab:results_13fix} summarizes the results of the combined fits which assume fixed $\theta_{13}$ to the reactor best-fit value~\cite{DayaBay:2018yms, The_Double_Chooz_Collaboration2020-bz, PhysRevLett.121.201801}. The data shows a strong preference for the normal mass ordering that will be assumed throughout the discussion in \Crefrange{sec:atm-params}{sec:dcp-result} and further assessed in \cref{sec:ordering}.

\subsection{Atmospheric parameters}\label{sec:atm-params}

The combination measures the atmospheric mass splitting and mixing angle with a precision competitive with accelerator experiments.
We obtain
\begin{align}
\dmsq = 2.44^{+0.06}_{-0.04}\times 10^{-3}\,\mathrm{eV}^2, &\,\,\,\,
\ssqtt = 0.52^{+0.03}_{-0.04}
\end{align}
assuming normal ordering.
\Cref{fig:headline} shows the two-dimensional confidence regions in the $(\ssqtt, \dmsq)$ plane for each experiment separately and for the combination.
The complementarity anticipated in Ref.~\cite{Arguelles:2022hrt} is realized in data: \DeepCore drives the precision on $\dmsq$ through its large statistics measurement of the oscillation minimum at $E_\nu\sim 20$\,\si{\GeV}, while \SK and ORCA6 are more sensitive to $\ssqtt$ through the appearance channel. The corresponding one-dimensional $\chi^2$ projections, obtained by profiling over the remaining parameters, are shown in \Cref{fig:results}.

The correlated treatment of the flux uncertainties tightens the combined $1\sigma$ interval by $9\%$ in $\ssqtt$ and by $19\%$ in $\dmsq$ relative to a trivial $\chi^2$ sum. The data is compatible with maximal mixing, with only a $\Delta\chi^2 = 0.3$ preference favoring the best fit in the upper octant.

\subsection{$CP$-violating phase}\label{sec:dcp-result}

The combination of atmospheric neutrino data provides a constraint on $\dCP$ that is independent and complementary to accelerator neutrino experiments. The sensitivity arises primarily from the Sub-GeV \SK samples, where the averaged solar-scale oscillations imprint a broad, $\dCP$-dependent distortion on the electron-like and muon-like distributions~\cite{Martinez-Soler:2019nhb,Kelly:2019itm}. This sensitivity is further enhanced by the neutron-tagged samples, which statistically separate neutrinos from antineutrinos.

\Cref{fig:results} shows the $\Delta\chi^2$ profile as a function of $\dCP$. We find a best-fit value of $\dCP = 1.30\pi^{+0.24\pi}_{-0.28\pi}$, with values around $\dCP = 0.3\pi$ disfavored at more than $2.7\sigma$.
This constraint is consistent with the T2K measurement~\cite{T2K:2021xwb} and with the joint T2K--NOvA fit~\cite{T2K:2025wet}, and provides an atmospheric-only constraint for the ongoing discussion of the tension between the two beam experiments~\cite{Jin:2026cpt}.

This measurement is the most sensitive to systematic uncertainties among the oscillation parameters considered in the analysis. The dominant contributions arise from uncertainties in the flux flavor and neutrino-antineutrino ratios, the CCQE cross sections, and the detector's ability to distinguish neutrinos from antineutrinos, namely the neutron production and tagging efficiency. ORCA6 and, to a lesser extent, IceCube provide larger statistics at $\mathcal{O}(1\,\mathrm{GeV})$ that help constrain some of the flux and cross-section uncertainties, yielding an enhancement of this measurement.

\begin{figure}
    \centering
    \includegraphics[width=\linewidth]{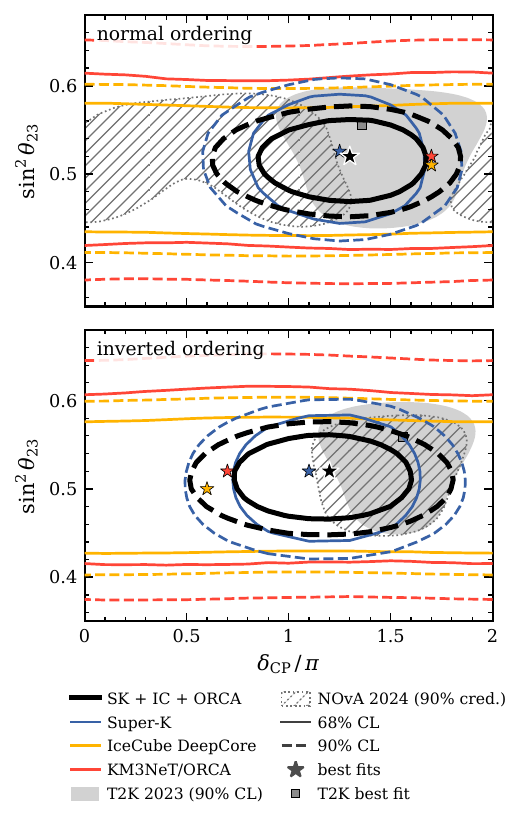}
    \caption{\textbf{\textit{Constraints in the $(\dCP,\ssqtt)$ plane.}}
            Allowed regions at 68\% (solid) and 90\% (dashed) CL for $(\dCP, \sin^2 \theta_{23})$ for NO and IO in the top and bottom panels respective. The combined analysis of \SK, \DeepCore and ORCA6 (thick black) is compared with our fits of \SK (blue), \DeepCore (orange) and ORCA6 (red) alone; stars mark the best-fit points. We also show for comparison the published 90\% regions of T2K~\cite{T2K:2023smv,T2K:2023data} and NOvA~\cite{NOvA:2025tmb,NOvA:2024data}  for the same ordering, each evaluated at its own best fit within that ordering.}
    \label{fig:dcp-s23}
\end{figure}

The constraints provided by atmospheric neutrino experiments can also be viewed in the $\sin^2\theta_{23}$--$\dCP$ plane, providing information complementary to that from long-baseline accelerator experiments. In particular, T2K~\cite{T2K:2023smv,T2K:2023data} and NOvA~\cite{NOvA:2025tmb,NOvA:2024data} exhibit a tension in this plane under normal ordering, while their preferred regions are more consistent under inverted ordering, shown in \cref{fig:dcp-s23}. This tension has important implications for the determination of the mass ordering, as the combination of the two experiments can modify the ordering preference obtained from the individual data sets. Independent information on both $\sin^2\theta_{23}$ and $\dCP$ is therefore particularly valuable for assessing the origin and impact of this tension.

The combination of atmospheric neutrino data provides complementary information in this parameter space. As discussed above, \SK provides most of the sensitivity to $\dCP$, while the large statistics of \DeepCore provide strong constraints on $\sin^2\theta_{23}$. The resulting constraints in the $\sin^2\theta_{23}$--$\dCP$ plane therefore provide an independent test of the regions preferred by accelerator experiments. As shown by the black contours in \cref{fig:dcp-s23}, the combined atmospheric data are in better agreement with the regions preferred by T2K for both mass orderings, while the precision on $\dCP$ still allows for compatibility with the NOvA.

\begin{figure*}
    \centering
    \includegraphics[width=0.95\linewidth]{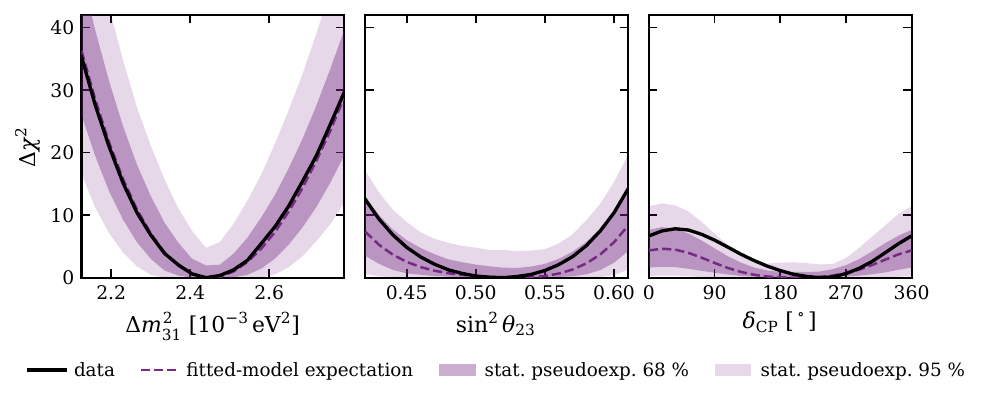}
    \caption{\textbf{\textit{Data fit $\Delta \chi^2$ and Asimov sensitivity with statistically fluctuated pseudoexperiments.}} For each of the physics parameters, 1000 pseudoexperiments were thrown with only Poisson-fluctuation around Asimov expectation at best fit physics and nuisance parameters, then fitted to produce the $68\%$ and $90\%$ containment bands for $\Delta \chi^2$, each computed with respect to the pseudoexperiment's own best fit point.}
    \label{fig:PE-bands}
\end{figure*}

\subsection{Neutrino mass ordering}\label{sec:ordering}

After profiling over all other parameters, we compare the normal- and inverted-ordering hypotheses and find a strong preference for the former. 
Comparing the best-fit points of the two orderings, we find
\begin{equation}
\chi^2_{\rm IO} - \chi^2_{\rm NO} = 11.0,
\end{equation}
Because the two orderings are non-nested, Wilks' theorem does not apply to determine the statistical significance from this difference. The caveats of interpreting such a comparison with Wilks' theorem are discussed in Ref.~\cite{Algeri:2019lah}. 

To assess the significance, we calibrate this test statistic with ensembles of pseudoexperiments generated under each ordering and fitted with the full analysis. We generate 1500 trials assuming normal ordering, for which the resulting $p$-value is relatively large, and 4397 trials assuming inverted ordering in order to fully capture the tail behavior of the test-statistic distribution. For IO, only 2 out of 4397 pseudoexperiments yield a test statistic larger than observed in the data, corresponding to 
\begin{equation}
    p_{\rm IO} = 2/4397 = 4.55\times10^{-4}
\end{equation}
This corresponds to a rejection of inverted ordering at $3.3\sigma$ (one-sided). For comparison, $p_{\rm NO}=0.915$ of the pseudoexperiments generated assuming normal ordering yield a test statistic less compatible with normal ordering than that observed in the data. The resulting distributions are shown in \Cref{fig:headline}.


The finite number of pseudoexperiments used to estimate the test statistics introduces a statistical uncertainty in the estimated $p$-values. Since each $p$-value is obtained from the fraction of pseudoexperiments falling in the corresponding tail, this uncertainty can be treated as a binomial uncertainty. We quantify it using the exact Clopper--Pearson (C-P) central interval at $68\%$ confidence~\cite{ClopperPearson1934}. For the inverted-ordering ensemble, with 2 tail events out of 4397 pseudoexperiments, this gives
\begin{equation}
p_{\rm IO} \in [1.62\times10^{-4},1.05\times10^{-3}].
\end{equation}
Converting this interval into a one-sided Gaussian significance using $Z=\Phi^{-1}(1-p)$, where $\Phi$ denotes the cumulative distribution function of the standard normal distribution and $\Phi^{-1}$ its inverse, gives
\begin{equation}
Z \in [3.08,3.60]\sigma.
\end{equation}
The central value corresponding to a significance of $3.32\sigma$.

For reference, we consider the modified frequentist $CL_s$ criterion~\cite{Junk:1999kv,Read:2002hq}, which accounts for the probability of obtaining a result at least as extreme as the observed one under inverted ordering relative to the sensitivity of the analysis to distinguish the two ordering hypotheses. We obtain
\begin{equation}
CL_s = \frac{p_{\rm IO}}{1-p_{\rm NO}}
= \frac{4.55\times10^{-4}}{1-0.915}
= 5.4\times10^{-3},
\end{equation}
corresponding to a one-sided Gaussian significance of $2.55\sigma$. The distributions of the test statistic for the two ordering hypotheses are shown in \Cref{fig:headline}.

The preference is driven by the matter-effect resonances: in the normal ordering, the MSW~\cite{Wolfenstein:1977ue,Mikheyev:1985zog} and parametric~\cite{Akhmedov:1988kd,Petcov:1998su} resonances enhance the conversion of neutrinos, rather than antineutrinos, crossing the mantle and core at 2--10\,\si{\GeV}, and the neutrino-dominated event rates of all three experiments respond coherently to this enhancement.
The pattern observed here, with each individual experiment showing a preference ($\Delta\chi^2 = 6.2$, 1.6 and 0.5 for \SK, \DeepCore and ORCA6) that strengthens to 11.0 in combination, matches the expectation from our forecast~\cite{Arguelles:2022hrt} and is consistent with the preferences reported by \SK~\cite{Super-Kamiokande:2023ahc} and by global fits~\cite{Esteban:2020cvm,deSalas:2020pgw}.



Additionally, we also discuss the strength of the data fit constraints on the physics parameters with respect to the Asimov sensitivity. \Cref{fig:PE-bands} shows the data fit $\Delta \chi^2$ curves on a band of pseudoexperiment results, where pseudoexperiments are thrown at $(\thetab_{\rm BF}, \etab_{\rm BF})$ with only Poisson statistical fluctuations. To interpret this plot, each subplot is effectively a spectrum of test statistic (TS) distributions: at every physics parameter point $\theta_0$, the $68\%$ and $90\%$ containment bands are plotted for the pseudoexperiment TS distribution where the data TS line is shown on top of it. For $\Delta m^2_{31}$, the data fit constraint is basically in agreement with Asimov everywhere; for $\sin^2\theta_{23}$, the fit strength is consistent within 1$\sigma$ of statistical fluctuations around Asimov expectation. For $\dCP$, the data fit exhibits exclusion strength above $68\%$ containment region for statistically fluctuated Asimov expectation, but mostly contained witin $90\%$. We include more discussions to nuisance parameter's effect on the fit in \cref{app:systs}.

\section{Joint Fit with JUNO}\label{sec:JUNO}

The Jiangmen Underground Neutrino Observatory (JUNO) is a 20-kton liquid-scintillator detector located at a baseline of about 52.5 km from the eight reactor cores of the Yangjiang and Taishan power plants, with a design energy resolution of about 3\% at 1 MeV~\cite{JUNO:2023ete}. 
At this baseline the reactor $\bar\nu_e$ survival probability carries both the slow solar oscillation driven by $\Delta m^2_{21}$ and the fast atmospheric oscillation driven by $\Delta m^2_{31}$, so that a single spectrum measures $\sin^2\theta_{12}$, $\Delta m^2_{21}$ and $\Delta m^2_{31}$ simultaneously. Since the $\Delta m^2_{31}$ and $\Delta m^2_{32}$ oscillation terms enter with different amplitudes, $\cos^2\theta_{12}$ and $\sin^2\theta_{12}$, the phase of the fast oscillation evolves differently across the reactor spectrum for normal and inverted ordering, making the spectral shape also sensitive to the mass ordering.

Recently, JUNO reported a measurement of the neutrino oscillation parameters based on 207 live days of data, achieving world-leading precision on the solar parameters, $\sin^2\theta_{12}$ and $\Delta m^2_{21}$, as well as a first determination of $\Delta m^2_{31}$ at the $\sim 1\%$ level~\cite{JUNO:2026tap}. With the current statistics, however, the fast oscillations are not yet sufficiently well resolved to determine a unique value of $\Delta m^2_{31}$: the $\chi^2$ profile exhibits several degenerate local minima. External information is therefore needed to select the value of $\Delta m^2_{31}$ consistent with other observations.


The combination of JUNO data with Daya Bay data and the constraint on $|\Delta m^2_{31}|$ from long-baseline (LBL) experiments yields a preference for normal mass ordering, with $\Delta\chi^2 = 4.86$, corresponding to a significance of $2.1\sigma$ after calibration using pseudoexperiments. As discussed above, atmospheric neutrino experiments provide a precise determination of $\Delta m^2_{31}$. Here, we present both a fit to JUNO data incorporating our combined atmospheric $|\Delta m^2_{31}|$ and a joint fit to the atmospheric neutrino and JUNO data. The latter is a straightforward extension of our atmospheric experiment combination, owing to the non-overlap of the systematic parameters.

\begin{figure}
    \centering
    \includegraphics[width=\linewidth]{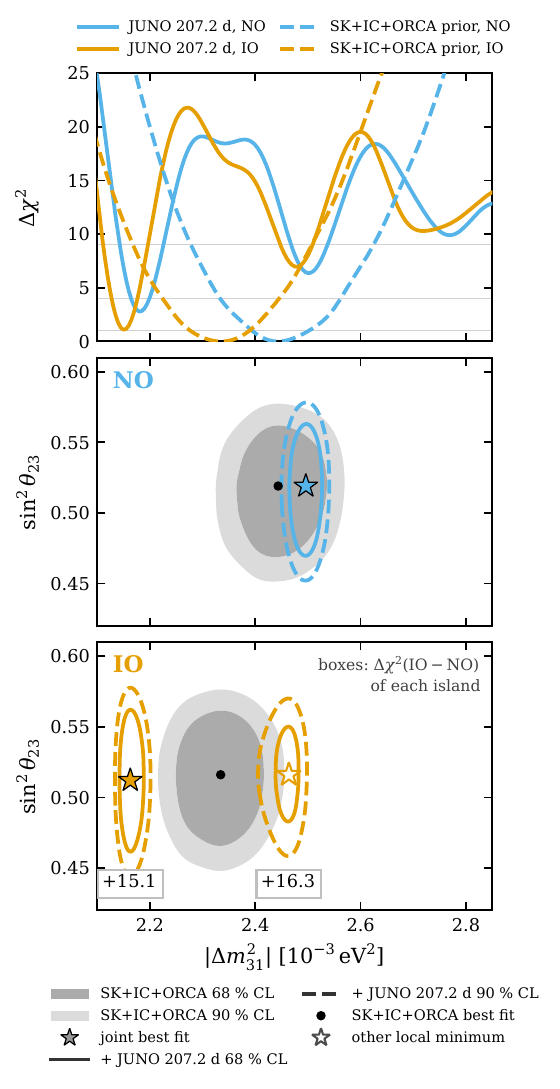}
    \caption{\textbf{\textit{Combination of the SK+IC+ORCA atmospheric fit with the JUNO 207.2-day reactor measurement.}} \textbf{Top}: the JUNO $\chi^2$ profile in $|\Delta m^2_{31}|$ under normal (blue) and inverted (orange) ordering, reconstructed from the published spectrum without any external $|\Delta m^2_{31}|$ constraint relative to its global minimum, together with our atmospheric $|\Delta m^2_{31}|$ profiles for each ordering (dashed), each relative to its own minimum.  \textbf{Middle and bottom}: allowed regions in the $(|\Delta m^2_{31}|, \sin^2\theta_{23})$ plane for NO and IO, profiled over $\delta_{\rm CP}$. The grey regions show the 68\% and 90\% C.L. allowed regions (2 d.o.f.) from the atmospheric fit alone, with the best-fit point marked by a dot. The solid and dashed contours show the corresponding regions after including the JUNO $\chi^2$ profile, with the joint best-fit points marked by stars. For IO, both allowed regions (islands) are shown, together with their respective best-fit points and the corresponding combined $\Delta\chi^2$ values.}
    \label{fig:juno-3-panel}
\end{figure}

We use the per-reactor-core spectrum from the Daya Bay 2025 unfolded antineutrino yield~\cite{DayaBay:2025ngb}, folded with a response built from the read-out non-linearity and resolution from JUNO experiment~\cite{JUNO:2026tap,2026junotao}, normalized to JUNO's own unoscillated curve. 
The TAO measurement~\cite{2026junotao} is not included in the results presented here; instead, we use only the constraint on the reactor flux from Daya Bay.
In the analysis, we include the predictions for the main background contributions provided by the JUNO collaboration, allowing their normalizations to vary independently. Further details of the implementation are provided in the \cref{app:juno}.

The results are shown in \cref{fig:juno-3-panel}. The top panel shows the JUNO reactor $\Delta m^2_{31}$ profiles from JUNO with the sequence of characteristic local minima, 
together with the $\Delta \chi^2$ profile from our combined atmospheric neutrino analysis, which does not include the sensitivity to the mass ordering. The best-fit values from the combination of atmospheric neutrino measurements lies at $2.44\times 10^{-3}\text{eV}^2$ for NO and $2.33\times 10^{-3}\text{eV}^2$ for IO (dashsed lines).
For NO, the atmospheric best fit lies close to one of the reactor minima, whereas for IO it falls between two of them. This mismatch under IO is what drives the additional sensitivity to the mass ordering when the reactor data are included.  
Using the JUNO profile alone, together with our combined atmospheric-fit profile for $|\Delta m^2_{31}|$ as a prior and a pull term, we obtain $\Delta \chi^2_{\rm NO-IO} = -4.10$, in favor of NO, which corresponds to a $2.01\sigma$ preference for NO once calibrated with pseudoexperiments.



The bottom two panels show the allowed regions in the $(|\Delta m^2_{31}|, \sin^2\theta_{23})$ plane from the joint atmospheric and JUNO analysis, for NO and IO. For NO, the combination shifts the preferred value of $|\Delta m^2_{31}|$ towards the minimum in the JUNO profile. For IO, the tension between the atmospheric and JUNO measurements of $|\Delta m^2_{31}|$ is clearly visible, with the combined analysis yielding two possible solutions that are in tension with the atmospheric data alone. This joint fit reaches $<1.0\%$ precision in the measurement of $|\Delta m^2_{31}|$ parameter, as shown in~\cref{tab:dm31_juno}. Using our combined atmospheric dataset, the IO solution at $|\Delta m^2_{31}| = 2.16 \times 10^{-3}\,\text{eV}^2$ is disfavored at $\Delta\chi^2_{\rm NMO} = 15.1$. The second minimum, which lies closer to the value of $|\Delta m^2_{31}|$ preferred under NO, is disfavored at $\Delta\chi^2_{\rm NMO} = 16.3$.

\begin{table}[t]
    \centering
       \begin{tabular}{c||c|c}
    Parameter & \makecell{SK+IC+ORCA} & \makecell{\textbf{SK+IC+ORCA}\\\textbf{+ JUNO 207 d}} \\\hline\hline
    \makecell{$\dmsq$ (NO)\\$[10^{-3}\,\mathrm{eV}^2]$} & $2.44^{+0.06}_{-0.04}$ & $2.50^{+0.02}_{-0.02}$ \\\hline
    \makecell{$\dmsq$ (IO)\\$[10^{-3}\,\mathrm{eV}^2]$} & $-2.33^{+0.05}_{-0.06}$ & $-2.16^{+0.01}_{-0.02}$ \\\hline
    $\Delta\chi^2(\mathrm{IO}-\mathrm{NO})$ & $11.0$ & $15.1$ \\
    \end{tabular}
    \caption{\textbf{\textit{Atmospheric mass splitting before and after adding JUNO.}}
    Best-fit values and $1\sigma$ intervals of $\dmsq$ from the combined SK+IC+ORCA fit (Table~\ref{tab:results_13fix}) and from its joint fit with the 207-day JUNO result, for normal and inverted ordering.In the joint inverted-ordering fit the minimum lies in the lower $|\dmsq|$ valley of the JUNO spectrum.}
    \label{tab:dm31_juno}
\end{table}

We assess the statistical significance of the mass-ordering preference by performing pseudoexperiments under both the NO and IO hypotheses. The results yield a preference for NO at a significance of $3.9\sigma$ in the combined atmospheric and JUNO analysis. A detailed description of the statistical analysis, including the generation and analysis of the pseudoexperiments, is provided in \cref{app:juno}. This appendix also presents the full implementation of our JUNO simulation and its validation against the official JUNO results~\cite{JUNO:2026tap,JUNO:2025first,2026junotao}.


\section{Conclusion}\label{sec:conclusions}
We have presented a measurement of the three-flavor oscillation parameters from a combined fit to real atmospheric neutrino data with correlated systematic uncertainties.
Using all data-taking stages of \SK with ultra-pure water, the latest \DeepCore data release, and ORCA6, we treat the shared atmospheric flux and neutrino-water cross-section uncertainties as common nuisance parameters across the three experiments, realizing in data the methodology we introduced as a forecast in Ref.~\cite{Arguelles:2022hrt}.

The combination delivers a complete picture of the measurement of the remaining neutrino oscillation parameters.
We measure atmospheric-sector precision competitive with accelerator experiments and independent of them, $\dmsq$ to $2.2\%$ and $\ssqtt$ to $6.1\%$, finding no preference for either octant (favors upper octant by $\Delta\chi^2 = 0.3$) with near-maximal mixing. Additionally, the data is able to constrain $\dCP$ without any beam input.
The correlated systematic treatment is not an ornament: it improves the constraint on $\ssqtt$ and $\dCP$ over a naive combination of published likelihoods, because the high-statistics samples of each experiment anchor the flux and cross-section parameters for all of them.

Our results add an independent witness to the question of the neutrino mass spectrum.
The combined atmospheric data find $3.32 \sigma$ evidence for the normal ordering, a preference that arises coherently from the Earth's matter effects in three detectors with different media, systematics, and analysis chains.
Each data set alone offers only a hint; together they point in one direction, and the upcoming data from the IceCube Upgrade, the full ORCA detector, SK-Gd, and JUNO will sharpen this picture within the decade~\cite{Arguelles:2022hrt,JUNO_ORCA}.
Like an outcome foretold long before it is pronounced, the verdict on the mass spectrum appears already written in the data: the normal ordering is looming.

The precise determination of $|\Delta m^2_{31}|$ from atmospheric neutrino experiments provides a complementary approach to determining the neutrino mass ordering when combined with reactor measurements~\cite{Blennow:2013vta,Blennow:2013oma,IceCube-Gen2:2019fet,KM3NeT:2021rkn}. Atmospheric and reactor experiments probe the atmospheric mass splitting through different disappearance channels, namely muon-neutrino and electron-antineutrino disappearance, respectively. As demonstrated in this work, the percent-level precision achieved by atmospheric measurements under both normal ordering (NO) and inverted ordering (IO) provides complementary information that can enhance the sensitivity to the mass ordering~\cite{Choubey:2003qx,deGouvea:2005hk,Nunokawa:2005nx}.


We have analyzed 207 days of JUNO data and combined the resulting constraints with our atmospheric-neutrino analysis. Using the JUNO $\chi^2$-profile together with the atmospheric constraint on $|\Delta m^2_{31}|$, we find a preference for normal ordering (NO) of $\Delta\chi^2 = 4.1$, corresponding to a significance of $2.0\sigma$ after calibration with pseudoexperiments. When the atmospheric and JUNO data are included in a joint fit, the preference for NO increases to $\Delta\chi^2_{\rm NMO} = 15.1$, corresponding to a significance of $3.9\sigma$. This improvement arises from the tension between the atmospheric and reactor constraints on $|\Delta m^2_{31}|$ under IO, illustrating the complementarity of the two experimental approaches and the potential of their combination to strengthen the determination of the neutrino mass ordering.

The normal ordering looms large over the neutrino community. Its determination is highly relevant for the next generation of long-baseline experiments, such as DUNE and Hyper-Kamiokande: it not only removes one of the unknowns these experiments aim to resolve, but also enhances their sensitivity to the remaining parameters, in particular the CP-violating phase. An independent determination of the mass ordering is especially important for experiments with smaller matter effects, such as Hyper-Kamiokande, since it breaks the degeneracy between the mass ordering and $\delta_{\rm CP}$~\cite{Hyper-Kamiokande:2018ofw, Hyper-Kamiokande:2025asb}, improving their prospects for establishing CP violation.





%
\begin{acknowledgments}
We thank Thomas Wester and the Super-Kamiokande Collaboration for discussions and help. CAA are supported by the Faculty of Arts and Sciences of Harvard University, Canadian Institute for Advanced Research (CIFAR), the National Science Foundation (NSF), the John Templeton Foundation, the Research Corporation for Science Advancement, and the David \& Lucile Packard Foundation.
PFM is supported by RYC2024-049537-I, funded by MICIU/AEI/10.13039/501100011033 and by ESF+.
IMS was supported by RYC2024-050360-I, funded by MCIU/AEI/10.13039/501100011033 and by ESF+, and by grant PID2025-172338NB-I00 funded by MICIU/AEI/10.13039/501100011033 and by ERDF,EU.
MJ is funded by the National Science Foundation (NSF) CAREER Award \#2239795.
\end{acknowledgments}

\section*{Code availability and AI Policy}
The central \texttt{PyNu} software is public at~\cite{pynu}.
The Monte Carlo simulation of SK can be downloaded from \cite{DVN/E1NVLA_2026}.

AI tools has been used in the development of this work in manuscript preparation, plot styling, code improvement and editing with human supervision and verification, based on central ideas and content developed by the authors.


\newpage
\bibliography{main}

\onecolumngrid

\newpage
\appendix

\newcounter{appendix}
\crefalias{section}{appendix}
\crefname{appendix}{Appendix}{Appendices}
\Crefname{appendix}{Appendix}{Appendices}

\section{Details of the Fit Engine and Super-Kamiokande Monte Carlo Modeling}\label{app:MC}
In this Appendix section we first discuss the fit engine following \cref{eq:chain} in more detail, but specifically in the context of the high-stats Monte Carlo simulation we develop for SK, and adjusting it to match the official data release. In the discussion of tuning the response kernel, we also briefly discuss the adjustments we made to the SK official data release to account for a previous bug through the help of Thomas Wester and the Super-Kamiokande collaboration.

Eq.~\ref{eq:chain} gives the forward computation of the expected counts per bin from a bare response kernel with null nuisance and oscillations effects. This is sufficient for sensitivity studies. However, in performing data fits, we need to match our nominal null predictions to the officially released values~\cite{super_kamiokande_collaboration_2023_8401262} to ensure best reproduction of published results. To do this, we rely on tune kernels $\ktune$, a linear map on the index space of $R$:

\begin{equation}
R'\;=\;\mathcal K R,\qquad
R'[i|e,k,E,Z]\;=\;\sum_{i',e',k',E',Z'}\mathcal K\big[i,e,k,E,Z\,\big|\,i',e',k',E',Z'\big]\;R[i'|e',k',E',Z'],
\label{eq:tunedef}
\end{equation}

This is an action we apply on the bare kernel before oscillations and nuisance. For each tune kernel, it acts on only a subset of the axes $\mathsf A$ (active axis) of $R$, and transforms the rest (passive axes $\mathsf B$) as an identity. Looking at the active axes within any passive cell, we find that 

\begin{equation}
\sum_{\mathsf A}(\mathcal K R)\;=\;\sum_{\mathsf A'}\Big[\sum_{\mathsf A}\mathcal K[\mathsf A|\mathsf A']\Big]R[\mathsf A'].
\end{equation}

Therefore there can be 2 types of tune kernels. Normalization-preserving tunes are ones where $\sum_{\mathsf A}\mathcal K[\mathsf A|\mathsf A']=1$ for every source index $\mathsf A'$ (a column-stochastic matrix along $\mathsf A$). Then $\sum_{\mathsf A}(\mathcal K R)=\sum_{\mathsf A}R$ in every passive cell: the tune only redistributes mass along its active
axes. These are migration tensors. Normalization-changing tunes are rescaling of overall expected event counts even at null hypothesis, where the simplest case is diagonal rescaling $\mathcal K=\mathrm{diag}(\lambda)$ with $\lambda$ a function of some index set. These are rescaling tensors.

More importantly, and tune kernel that we might want to apply factorizes cleanly as (normalization-changing-diagonal) $\circ$ (migration).

To specify a tune kernel, we need to find a target to tune toward (an official release document) and an anchor point (the released bin count numbers). The normalization-changing, rescaling tunes are with respect to a release table, where the anchor point is at a particular physics parameter point usually with null nuisance $(\vec \theta^\ast, \etab)$, so we can compare our bare kernel to it at each passive cell of this rescaling tune $\mathsf P$ using

\begin{equation}
\lambda_{\mathsf P}\;=\;\frac{\text{Target}_{\mathsf P}}
{\sum_{\text{active}}R\,\phi(\vec \theta^\ast)\big|_{\mathsf P}}
\label{eq:pin}
\end{equation}

Notice that this only fixed the prediction at one point to a given nominal value or table, and away from this anchor point, the tuned kernel $\ktune R$ still follows the bare kernel's own $\vec \theta$-dependence. This is commonly applied once to match the reconstructed analysis bin contents (integrated over MC simulated true space distributions) at either the nominal or best fit physics parameters, at nominal nuisance values.

\begin{figure}
    \centering
    \includegraphics[width=\linewidth]{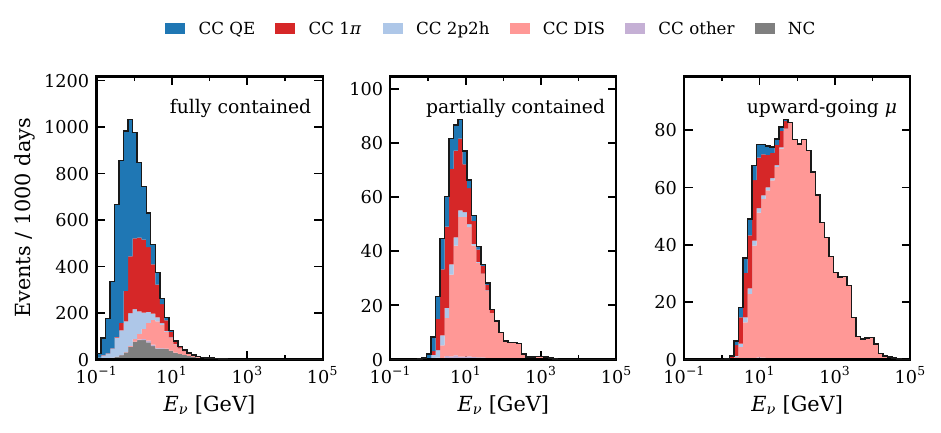}
    \caption{True energy distribution of high-statistics SK MC we develop after applying tune tensors to match for the SK official data release information on reconstructed bin counts and true energy distribution quantiles in the bins. The colors are the interaction modes composition in each true energy bin for each sample.}
    \label{fig:SK-Etrue-dist}
\end{figure}

The bare kernel $R$ is simply produced by binning over our high-statistics MC for SK, then we apply two such normalization tune kernels. The first one tunes the reconstructed analysis bin contents (integrated over true MC energy) to match the official release exactly, where the target we choose is the SK official release reconstructed bin counts at the reported NO best fit parameters with nominal nuisance~\cite{Super-Kamiokande:2023ahc,super_kamiokande_collaboration_2023_8401262}. The second tune targets the true energy bin counts within each reconstructed analysis bin, where we normalize to match the reported quantiles of the true energy distributions. These tunes are artificial reweights to match the reported information, and therefor leaves an imprint on the unoscillated true energy distributions of the samples when the reconstructed bins are summed over. However, even with some introduced kinks, the tuned response tensor $\mathcal K R$ still agrees well with SK Collaboration results~\cite{Super-Kamiokande:2023ahc}, as shown in \cref{fig:SK-Etrue-dist}.

The relative fractions of each interaction mode in the energy distribution shown in \cref{fig:SK-Etrue-dist} points to another type of possible tune tensor: migration tensors.

Migration tensors are defined by comparing the bare fractions in some active axes with the target fractions. Here we use the example of the cross section modeling. In the SK analysis, the generator for cross section is NEUT~\cite{Hayato:2021heg}, which is proprietary software; therefore, in our high statistics MC simulation for SK, we needed to replace it with GENIE~\cite{Andreopoulos:2009rq}. Supposing we do have access to NEUT cross section simulation that's also event-selection-aware, we can in principle tune the bare kernel to match the different modeling. We use this hypothetical situation to explain the other type of tune kernel: migration tunes.

These are simply migration matrices that are normalization-preserving, and act only on some active axis (the ''mode" bins) while not affecting the other axes (passive axes) of the bare response

\begin{equation}
R_{\NEUT}[i|e,(f,m'),E,Z]\;=\;\sum_{m}M_{m'm}\big(i,e,f,E,Z\big)\,R_{\GENIE}[i|e,(f,m),E,Z],
\label{eq:migration}
\end{equation}

Here the flavor $f$ is held fixed for each migration. The mode composition shuffling is hereby a migration only due to the conservation of total event count in each passive axis:

\begin{equation}
\sum_{m'}M_{m'm}=1\ \ \forall m
\;\;\Rightarrow\;\;\sum_{m'}(MR)[\cdots m'\cdots]=\sum_m R[\cdots m\cdots],
\label{eq:Mmass}
\end{equation}

This is to say, $M$ is block-diagonal in the indices $(i,e,f,E,Z)$, and commutes with every tensor that is constant along the mode bins. 

\section{Details of the Data Fit Results}\label{app:fit}
\begin{figure}[p]
    \centering
    \includegraphics[width=0.97\linewidth]{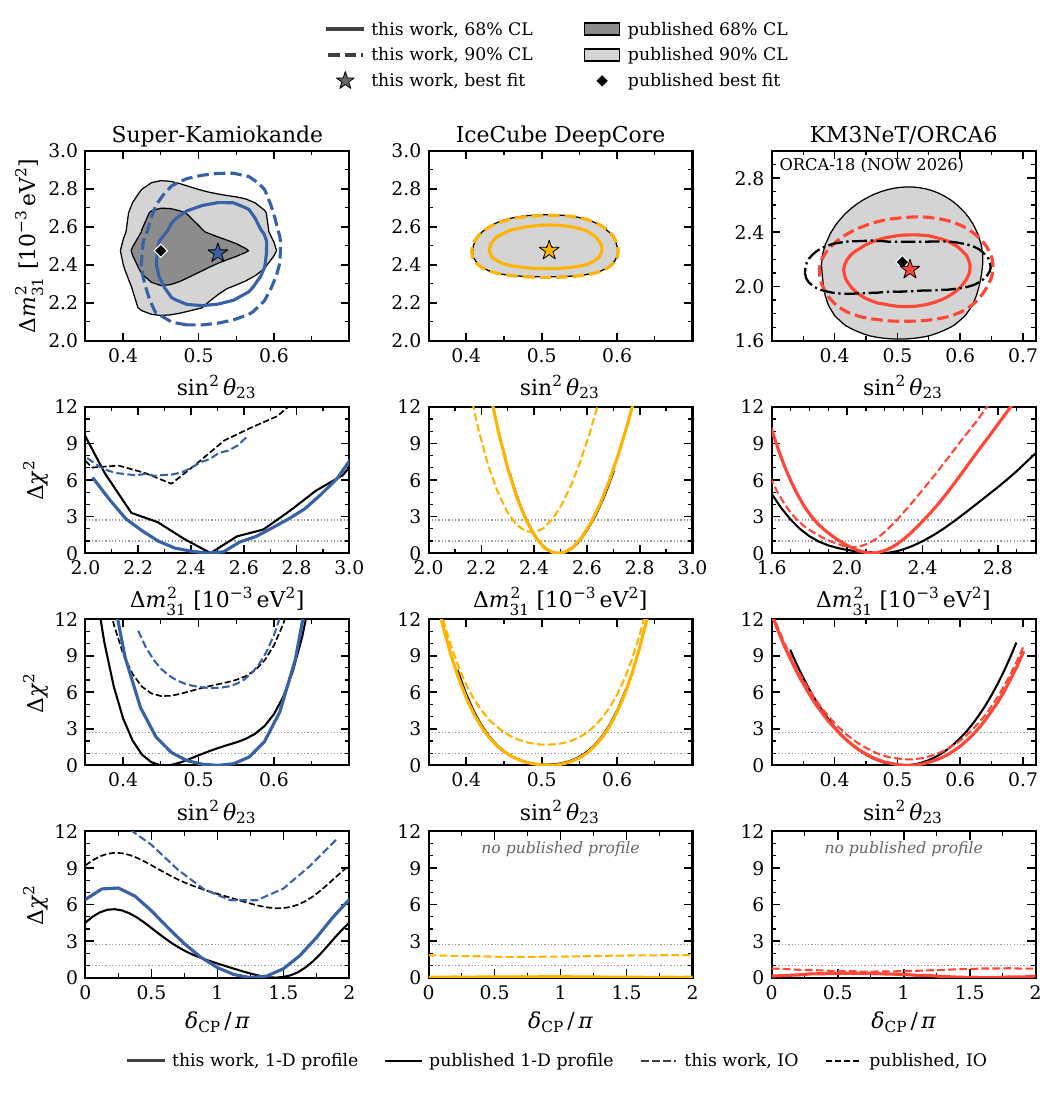}
    \caption{\textbf{\textit{Single-experiment fits compared with the published results.}}
    We show each experiment's public release contours and profiles against the fit results we obtain from our fit. In the contours, we show the $68\%$ and $90\%$ regions. Specifically for ORCA-6, we also overlay a recently published ORCA-18 contour~\cite{KM3NeT:ORCA18NOW2026}. 
    No official $\dCP$ profile information is available for \DeepCore or ORCA-6 for comparison. Values published in $\Delta m^2_{32}$ (\SK, \DeepCore) are converted with $\dmsq = \Delta m^2_{32} + \Delta m^2_{21}$, $\Delta m^2_{21} = 7.41\times10^{-5}$\,eV$^2$; all fits use $\sin^2\theta_{13} = 0.022$.}
    \label{fig:solo-vs-release}
\end{figure}

In this Appendix, we provide details to the data fit results. To begin with, we elaborate on the agreement of our single-experiment fits against the officially released contours and profiles in \cref{fig:solo-vs-release}. These fits are done with the same forward model as described in~\cref{eq:chain} with the same nuisance parameters as in the combined fit shown in the main text, except on a single experiment at a time. For \SK, we use our independently developed event-by-event MC, whereas we use the binned release provided by the IceCube and ORCA Collaborations~\cite{IceCubeCollaboration:2023wtb,km3netcollaboration2024measurementneutrinooscillationparameters}.

For SK, we see that the contours show a very mild ($< 1\sigma$) discrepancy in the best fit location, mainly due to the profile difference in $\sin ^2 \theta_{23}$, where we find the best fit value to be in the upper octant, closer to the maximal mixing. We see agreement in $\Delta m^2_{31}$ where we reproduce the quadratic behavior.

For IceCube, we use the detector response and systematics uncertainties provided in data release~\cite{IceCubeCollaboration:2023wtb}. In addition, we implement the flux and cross-section systematics uncertainties described in Section~\ref{sec:analysis}. Our results are in near-perfect agreement with the official release.

For ORCA-6, we find good agreement in $\sin^2 \theta_{\rm{23}}$, but find a narrower profile for $\Delta m^2_{31}$. This is due to the missing energy shift nuisance parameter, which is included in the ORCA official fit; it is however not modeled in the data release which includes only binned true and reconstructed energy information, whereas the energy shift parameter is a detector-response-level nuisance. This does not induce any significant impact on the combined fit. For reference, we also included a recent result~\cite{KM3NeT:ORCA18NOW2026} that shows a more livetime fit from the KM3NeT/ORCA Collaboration which already reaches a better constraint on the $\Delta m^2$ direction.

Additionally, we provide the details on individual experiments' goodness-of-fit analysis to support the parameter goodness-of-fit test statistic we provided in the main text. The goodness of fit estimated from 1000 pseudoexperiments generated at the combined best fit under normal ordering for each of the individual experiments.
For each pseudoexperiment every constrained nuisance parameter is drawn from a Gaussian of its prior width centered on its best-fit value, with parameters shared between experiments drawn once, and the expected counts in all analysis bins are Poisson-fluctuated; each pseudoexperiment is then fitted with the full analysis chain, where minimization is performed on all physics as well as nuisance parameters.

\begin{figure}[htbp]
    \centering
    \includegraphics[width=\linewidth]{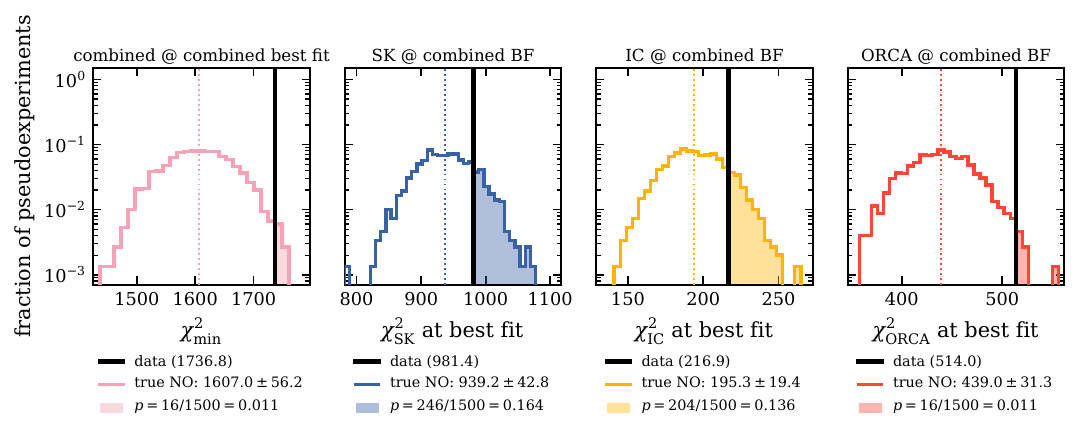}
    \caption{\textbf{\textit{Goodness of fit of the combined analysis.}}
    Distributions over 1500 pseudoexperiments generated and fitted under normal ordering (colored) compared with the data (black line).
    Shaded areas mark the fraction of pseudoexperiments at or above the data; dotted lines mark the medians.}
    \label{fig:gof}
\end{figure}

\Cref{fig:gof} shows the distribution of the minimum $\chi^2$ and of its three experimental contributions at the best fit.
The observed $\chi^2_{\rm min} = 1736.8$ is exceeded by  16 of the 1500 pseudoexperiments ($p = 0.011$).
The contributions of \SK and \DeepCore are typical ($p = 0.16$ and $0.14$), while that of ORCA6 lies in the upper tail ($p = 0.011$).

These p-values and $\chi^2$ test statistics reported above are evaluated using the Poisson likelihood without accounting for finite MC statistical uncertainties. This approximation is almost exact for \SK, thanks to the large MC sample we have generated. For \DeepCore and ORCA6, however, accounting for finite MC statistics in addition to Poisson fluctuations when generating pseudoexperiments at the best-fit point leads to less extreme $\chi^2$ test statistics.

\begin{figure}[htbp]
    \centering
    \includegraphics[width=\linewidth]{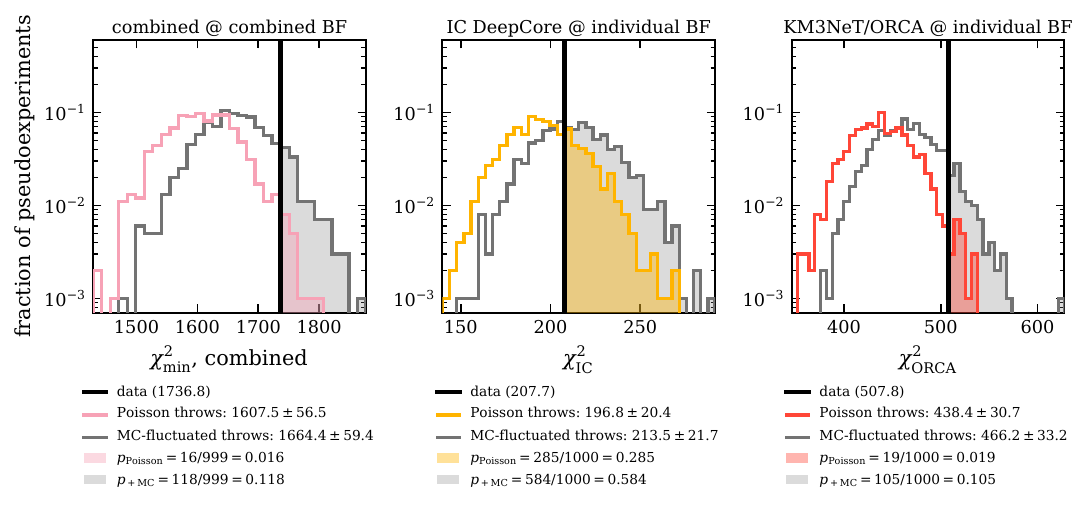}
    \caption{\textbf{\textit{Effect of the finite Monte Carlo statistics of the public releases.}}
    Goodness of fit of \DeepCore and ORCA6 fitted alone at their best-fit oscillation parameters, with all nuisance parameters profiled but oscillation parameters fixed; as well as the combined fit when throwing IceCube and ORCA realizations with the finite-MC and statistical fluctuations.}
    \label{fig:solo-gof-mc}
\end{figure}

This effect is illustrated in \Cref{fig:solo-gof-mc}, where we fit \DeepCore and ORCA6 independently and compare the resulting test-statistic distributions with and without the finite-MC-statistics treatment. For efficiency, since we are only comparing the relative $p$-values with and without finite MC statistics uncertainties, we only profile the nuisance instead of the oscillations parameters too when fitting each pseudoexperiment realization. Including these uncertainties increases the goodness-of-fit p-values for both the individual experiments and the combined analysis. The combined goodness-of-fit then exhibits only mild tension, with (p=0.12), corresponding to a one-sided Gaussian significance of $1.18\sigma$.


\begin{figure}[p]
    \centering
    \includegraphics[width=\linewidth]{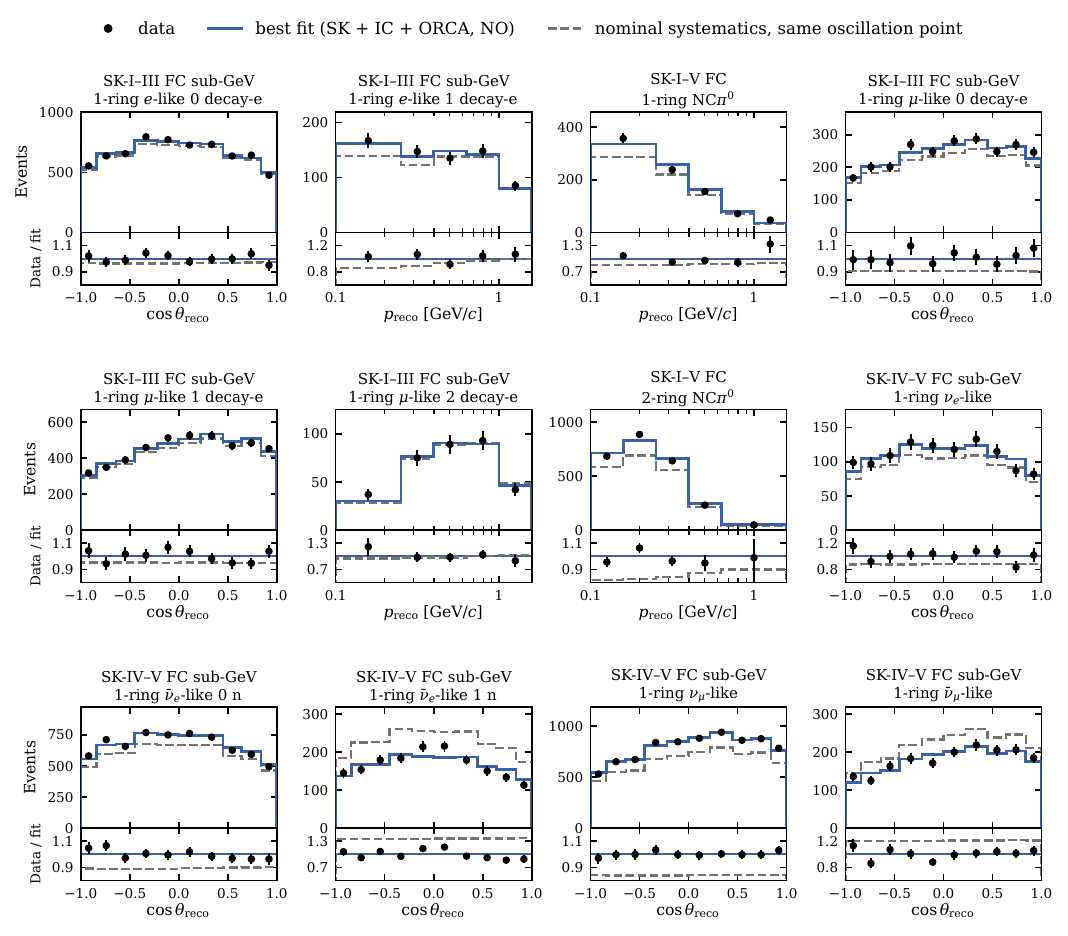}
    \caption{\textbf{\textit{\SK fully contained sub-GeV samples at the best fit.}}
    Observed event distributions of the 12 fully contained sub-GeV \SK samples (points, statistical uncertainties) compared with the prediction at the combined best fit in normal ordering (solid) and with the prediction at the same oscillation parameters and nominal systematic parameters (dashed).
    Zenith-binned samples are shown in reconstructed $\cos\theta$ summed over momentum, samples with a single zenith bin in reconstructed momentum; the lower panels show the ratio of data to the best fit.}
    \label{fig:bestfit-sk-1}
\end{figure}

\begin{figure}[p]
    \centering
    \includegraphics[width=\linewidth]{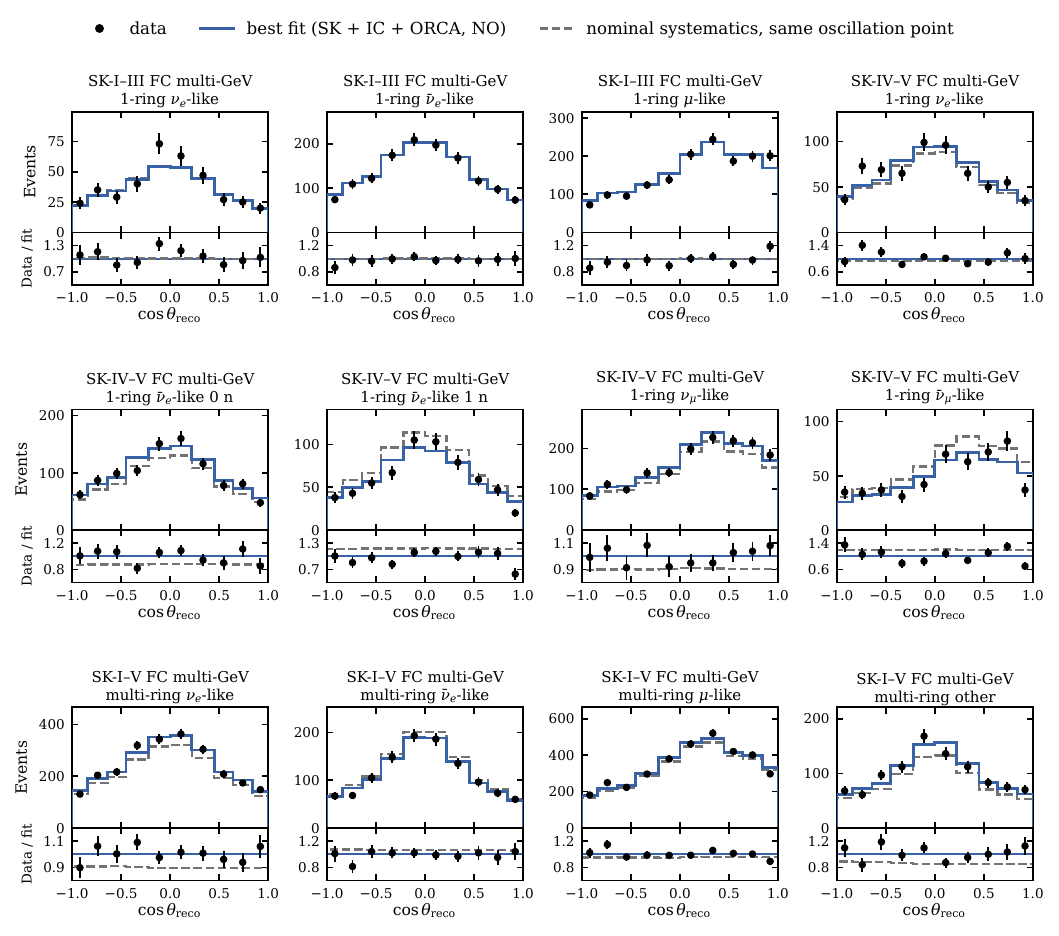}
    \caption{\textbf{\textit{\SK fully contained multi-GeV samples at the best fit.}}
    As \cref{fig:bestfit-sk-1}, for the 12 fully contained multi-GeV single-ring and multi-ring samples.}
    \label{fig:bestfit-sk-2}
\end{figure}

\begin{figure}[p]
    \centering
    \includegraphics[width=\linewidth]{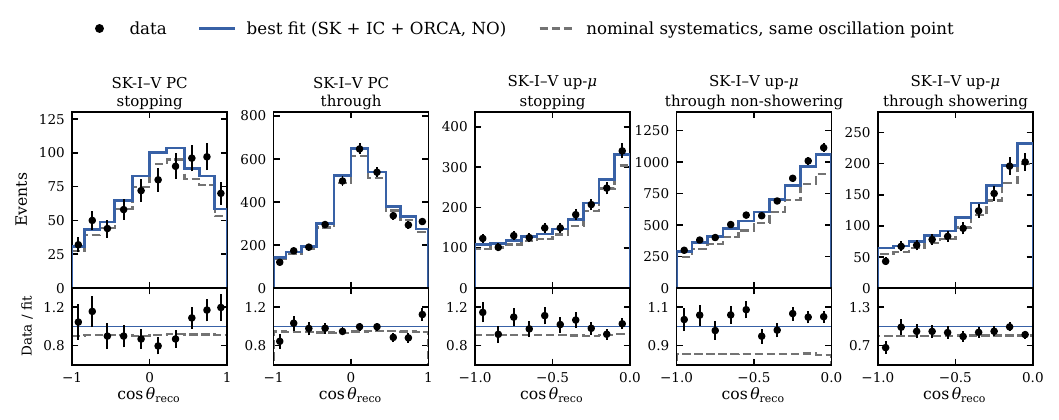}
    \caption{\textbf{\textit{\SK partially contained and up-going muon samples at the best fit.}}
    As \cref{fig:bestfit-sk-1}, for the two partially contained and the three up-going muon samples; the up-going muon samples cover $\cos\theta\in[-1,0]$.}
    \label{fig:bestfit-sk-3}
\end{figure}

\begin{figure}[p]
    \centering
    \includegraphics[width=\linewidth]{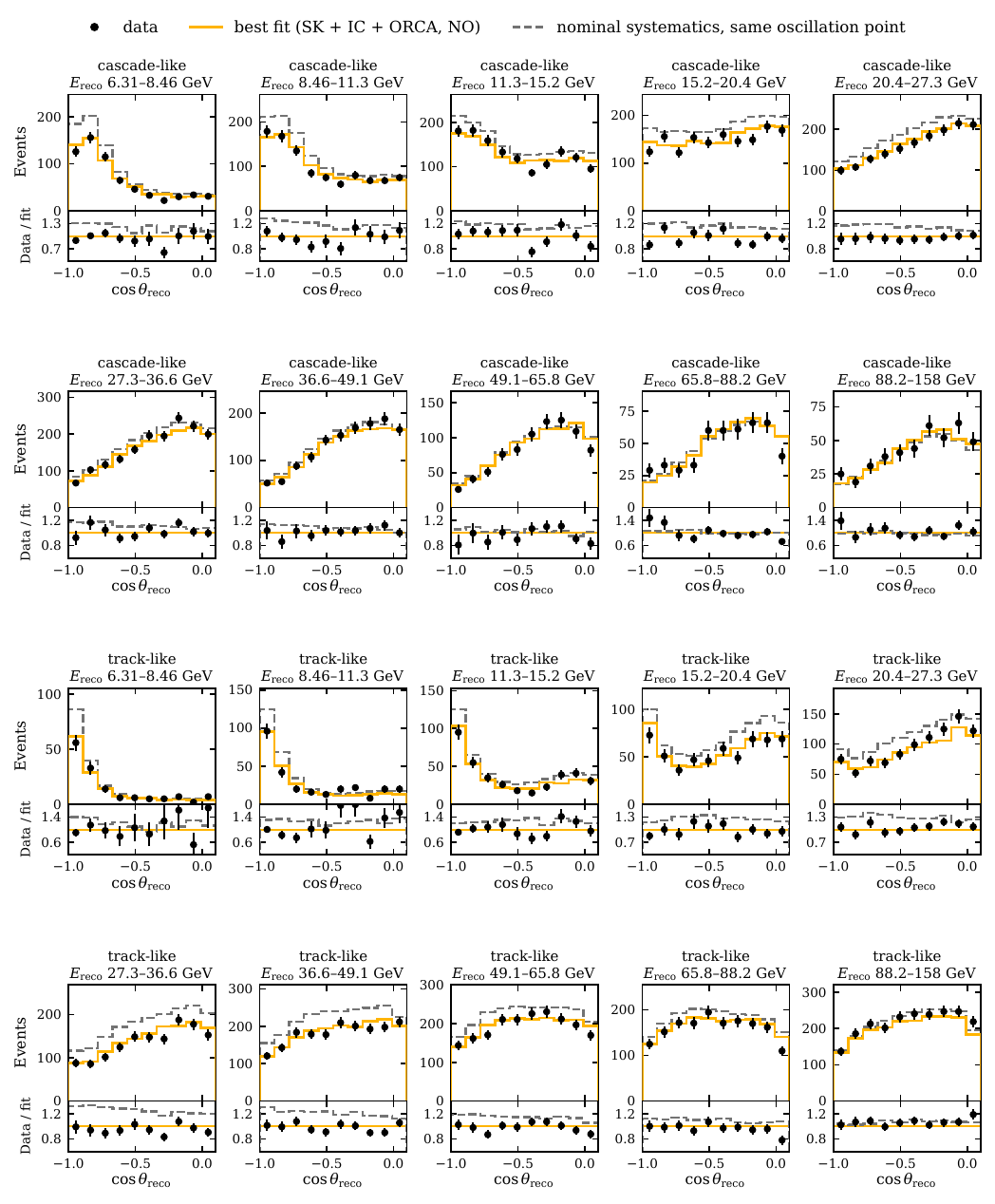}
    \caption{\textbf{\textit{\DeepCore event distributions at the best fit.}}
    Observed reconstructed $\cos\theta$ distributions in each of the ten reconstructed-energy slices, for the cascade-like (top two rows) and track-like (bottom two rows) event classes, compared with the combined best fit in normal ordering (solid) and the nominal-systematics prediction at the same oscillation parameters (dashed), with the ratio of data to the best fit below each panel.}
    \label{fig:bestfit-ic}
\end{figure}

\begin{figure}[p]
    \centering
    \includegraphics[width=\linewidth]{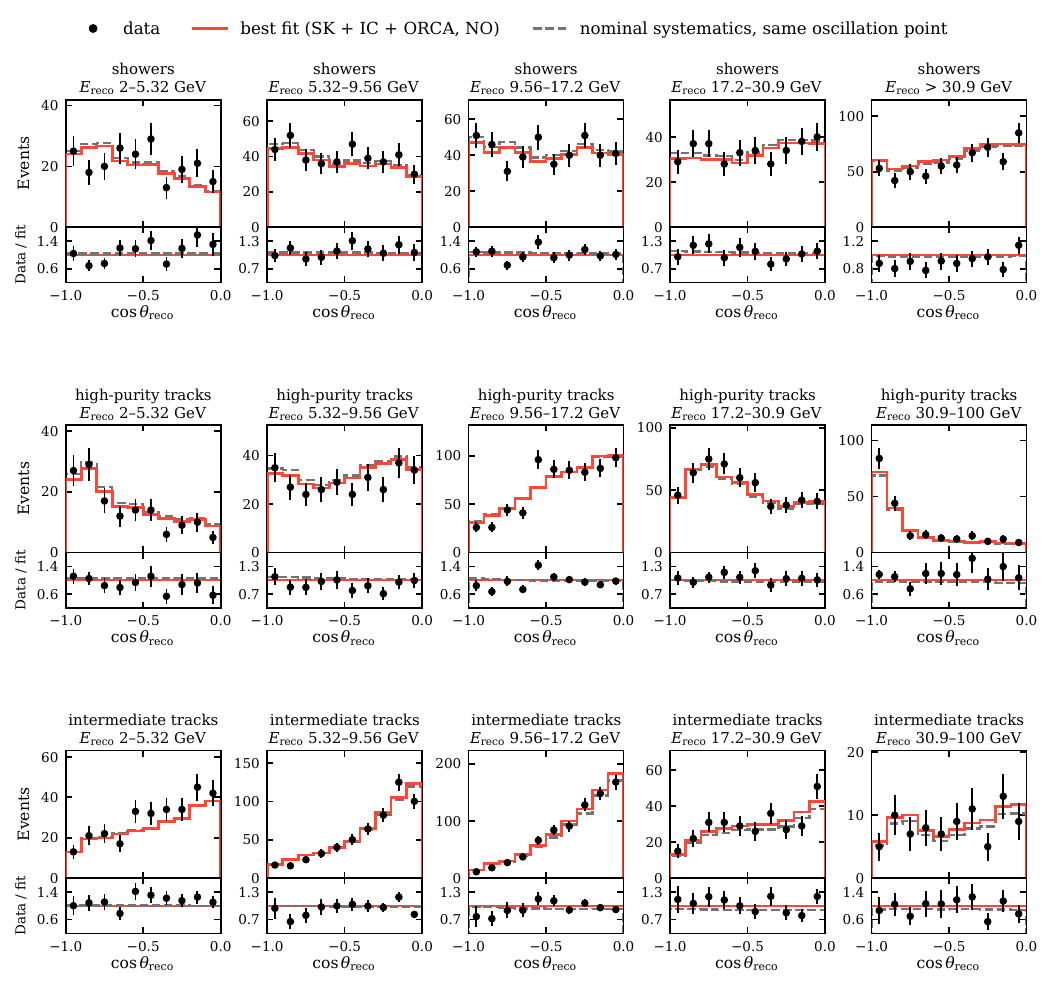}
    \caption{\textbf{\textit{ORCA6 event distributions at the best fit.}}
    Observed up-going reconstructed $\cos\theta$ distributions for showers, high-purity tracks and intermediate tracks (rows), each panel merging three of the released energy slices, compared with the combined best fit in normal ordering (solid) and the nominal-systematics prediction at the same oscillation parameters (dashed), with the ratio of data to the best fit below each panel. The atmospheric-muon background is included in the prediction.}
    \label{fig:bestfit-orca}
\end{figure}

Finally, we show the event distributions of every analysis sample in the 3 experiments at the combined best fit parameters $\dmsq = 2.44\times10^{-3}$\,eV$^2$, $\ssqtt = 0.52$ and $\dCP = 1.30\pi$ in normal ordering in \Cref{fig:bestfit-sk-1,fig:bestfit-sk-2,fig:bestfit-sk-3,fig:bestfit-ic,fig:bestfit-orca}, attached at the end of the Supplemental Materials.

\section{Systematic Uncertainties and Impact on the Fit}\label{app:systs}
In this Appendix section we provide information on the systematics parameter's impact to the data fit, as well as discuss the interpretation of the nuisance pulls in the context of Asimov signal and data-expectation residuals. 

\begin{figure}[h]
    \centering
    \includegraphics[width=\linewidth]{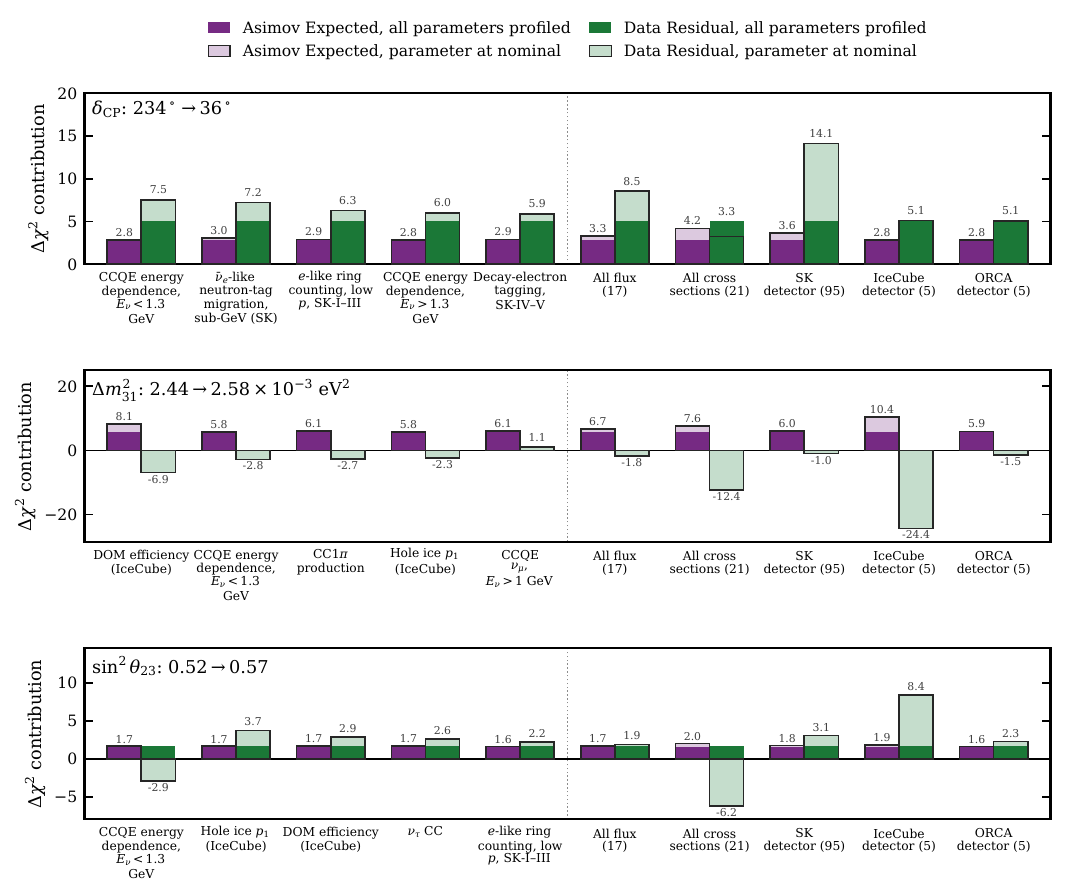}
    \caption{\textbf{\textit{$\Delta \chi^2$ decomposition for physics parameters and impact by nuisance parameters.}} For each depth (rows), the Asimov-expectation side and the data-residual terms between the best-fit point and the far point, with all 143 nuisance parameters profiled (solid) and with the named parameter or group held at nominal while all others are profiled . A bar outline above the solid bar means the parameter was reducing that side of the depth; an outline inside the solid bar means the profile is shallower without it; a negative value means the far point is preferred on that side. Each row shows the five individual parameters that change that depth most when held at nominal, followed, right of the dotted line, by the five parameter families. Top row: $\dCP$ $234^\circ\to36^\circ$; middle row: $\dmsq$ $2.44\to2.58\times10^{-3}$\,eV$^2$; bottom row: $\ssqtt$ $0.52\to0.57$.}
    \label{fig:app-e9c}
\end{figure}

To begin with, we provide a complete list of the shared and experiment-specific nuisance parameters in \cref{tab:systs}, included at the end of the Appendix sections. This table reports the nominal value, the $1\sigma$ region, and the pull at the combined best-fit point for each nuisance parameter. We observe that of the 143 nuisance parameters, most remain within their $1\sigma$ prior range, with seven pulled beyond $1\sigma$ (highlighted in light gray) and only one pulling beyond $2\sigma$, corresponding to the CCQE normalization shared by \DeepCore and \SK.

When studying the nuisance parameters effect on a fit to data, rather than on a sensitivity study, it can be useful to separate their effects on the Asimov prediction from their effects on the residuals between the data and the best-fit expectation. The former describes how well a nuisance parameter can absorb the changes in the expected event distribution brought by shifting a physics parameter, for example, the $\dCP$ distortion that appears in the sub-GeV energy bins. The latter describes how the same nuisance parameter can account for the fluctuation and mismodeling that we observe in data relative to the best-fit expectation. 

To study the effect of systematic uncertainties on $\Delta \chi^2$, we choose a common Asimov reference point at $(\thetab, \etab_{\rm nominal})$, where the physics parameters are fixed to their best-fit values obtained from data and the nuisances are set to their nominal values.  
Since we are studying the $\Delta \chi^2$, absolute shifts in the Asimov will enter both terms in the difference and therefore cancel. This defines the Asimov bin counts to be 
\begin{equation}
    O^{\rm A}_i\;\equiv\;E_i(\theta_{\rm BF},\vec\eta_{\rm nom}),
\end{equation}

Therefore at any physics parameter point $\thetab$, the $\chi^2$ distance to the Asimov can be computed as
\begin{equation}
    \mathcal A(\vec\theta) = \min\limits_{\etab} \chi^2_{\rm A}(\thetab) = \min\limits_{\etab} \chi^2(\thetab; O_{\rm{A}}) = d(O | E(\thetab, \etab)) + \Pi(\etab)
\end{equation}
where $d$ is the Poisson distance and $\Pi$ is the nuisance penalty term, as read off of \cref{eq:chi2-poisson}.

Simply then, the residual term is the data-derived constraint deviation from the Asimov. This residual can originate from several sources, including statistical fluctuations and mismodeling.  We can then write it simply as 
\begin{equation}
    \mathcal C(\vec \theta) = \chi^2(\thetab; O_{\rm{data}}) - \mathcal A(\thetab)
\end{equation}

For any physics parameter $\thetab_{\rm far}$, we can evaluate $\Delta \chi^2(\thetab_{\rm{far}}; \thetab_{\rm BF})$ of a far point with respect to the best fit point we obtain from data. The $\Delta \chi^2$ can be decomposed into an Asimov and a residual component as  
\begin{equation}
    \Delta\chi^2\;=\; \Delta \mathcal{A} + \Delta \mathcal{C} =\big(\mathcal A_{\rm far}-\mathcal A_{\rm BF}\big)+\big(\mathcal C_{\rm far}-\mathcal C_{\rm BF}\big)
\end{equation}

In order to study the effect of a nuisance parameter on the fit then, we can evaluate the decomposition of the $\Delta \chi^2$ value on best fit nuisance parameters versus on nominal nuisance parameters. To assess the effect of a single nuisance parameter $\eta_0$ in its ability to absorb Asimov signal and data fluctuation and mis-modeling signals, we compute the two deviations $\Delta \mathcal{A}_{\cancel{\eta_0}}$ and $\Delta \mathcal{C}_{\cancel{\eta_0}}$, where we minimize over all nuisance but $\eta_0$. Comparing this quantities with the corresponding fully profiled values provides a measure of the extent to which the nuisance parameter $\eta_0$ can absorb the constraining power arising from the Asimov prediction and from the data residual.  
%
\begin{align}
    \Delta \mathcal{A}_{\cancel{\eta_0}} &= A^{\rm far}_{\cancel{\eta_0}}-\mathcal A^{\rm BF}_{\cancel{\eta_0}} = \min\limits_{\etab \backslash \eta_0} \chi^2_{\rm A}(\thetab_{\rm far}) - \min\limits_{\etab \backslash \eta_0} \chi^2_{\rm A}(\thetab_{\rm BF}); \\
    \Delta \mathcal C_{\cancel{\eta_0}} &= \Big( \min\limits_{\etab \backslash \eta_0} \chi^2(\thetab_{\rm far}; O_{\rm{data}}) - A^{\rm far}_{\cancel{\eta_0}}\Big) - \Big( \min\limits_{\etab \backslash \eta_0} \chi^2(\thetab_{\rm BF}; O_{\rm{data}}) - A^{\rm BF}_{\cancel{\eta_0}}\Big)
\end{align}

Each panel of \cref{fig:app-e9c} shows this comparison between the pairs $(\Delta \mathcal{A}_{\cancel{\eta_0}}, \Delta \mathcal{A})$ and $(\Delta \mathcal{C}_{\cancel{\eta_0}}, \Delta \mathcal{C})$ for the five nuisance parameters with the largest pulls, together with the rest of the nuisance grouped into flux, cross-section, and respective detector systematics. Comparing $\Delta \mathcal{A}$ and $\Delta \mathcal C$ (darker colors, which are the same height across all nuisance ticks of each horizontal slice) tells us about how much sensitivity for each physics parameter is expected from Asimov alone, and how much is brought about by data-Asimov residuals, statistically or mis-modelling-induced. Comparing $\Delta \mathcal{A}_{\cancel{\eta_0}}$with$ \Delta \mathcal{A}$ (or the same with $\mathcal C$), we see the effect of each nuisance at absorbing the constraint power in Asimov and in data-induced residuals.

The different panels show the impact on different physics parameters.
For $\dCP$, we pick the point that correspond to the most excluded value. For $\Delta m^2$ and $\sin^2\theta_{23}$, we pick a moderately excluded high mass splitting and mixing angle, since both $\Delta \chi^2$ curves are quadratically shaped.

The $\Delta m^2_{31}$ panel provides a clear picture: the data-Asimov residual is almost cleanly absorbed by the nuisance parameters, and our power in measuring $\Delta m^2_{31}$ is almost completely expected, as we would see in an Asimov sensitivity study. For $\sin^2 \theta_{23}$, we see equal contribution in the constraining power in expected Asimov and residual data fluctuations, which pulls sensitivity on both sides, with nuisance parameters absorbing a lot of the residual sensitivity. In contrast, the constraining power on $\dCP$ leans towards the residual side, gaining more from data fluctuations around the Asimov expectation.

\section{Details on JUNO Simulation and Joint Fit}\label{app:juno}
\begin{figure}
    \centering
    \includegraphics[width=\linewidth]{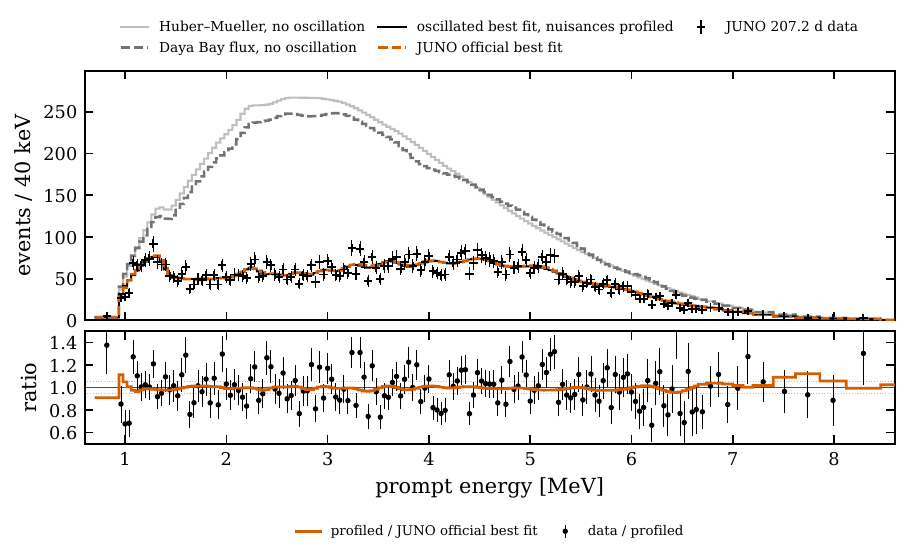}
    \caption{\textbf{\textit{Unoscillated and oscillated fluxes by this work and compared to the results in ~\cite{JUNO:2026tap}.}} \textbf{Top}: the 8294 inverse-beta-decay
    candidates read out from the published figure (crosses, statistical
    uncertainties), our prediction without oscillations for the Huber--Mueller
    flux (light grey) and for the Daya Bay unfolded flux used in this work
    (dark grey, dashed), our oscillated prediction at the JUNO best-fit
    oscillation parameters with the 34 nuisance parameters profiled (black),
    and the JUNO best-fit curve digitised from the same figure (orange,
    dashed).
    \textbf{Bottom}: the ratio of the data to our profiled prediction (points) and of
    our profiled prediction to the JUNO best-fit curve (line).}
    \label{fig:junofig3}
\end{figure}

In this Appendix we discuss the details of the implementation of our JUNO simulation.

JUNO has published two data sets: the first measurement with 59.1 days of exposure, for which the collaboration released the reconstructed spectrum, background histograms, detector-response curves and oscillation results as source data~\cite{JUNO:2025first}, and the 207.2-day analysis~\cite{JUNO:2026tap}, for which no official data release exists at the time of writing, and we recover every input from the vector graphics of the arXiv manuscript.

\textbf{\textit{Reactor flux and oscillation.}} We use the unfolded Huber-Muller spectra as a template of the unoscillated antineutrino spectrum, which is then tuned to match the Daya Bay measured total yield per fission~\cite{DayaBay:2025ngb}. The reactor configuration comprises the eight Yangjiang and Taishan cores at 52.1--52.8,km and the Daya Bay complex at 215 km. 

The survival probability is evaluated core by core in constant-density matter with $\rho = 2.55$ g/cm$^3$~\cite{JUNO:2025first}, on a 7000-grid in true energy, with $\sin^2\theta_{12}$, $\Delta m^2_{21}$ and $\Delta m^2_{31}$ free and $\sin^2\theta_{13}$ fixed to the Daya Bay value. 

We apply the computed oscillation probabilities onto the unfolded true neutrino energy spectra, and fold in the detector response to recover the reported event distribution in reconstructed prompt energy. In this process, we separate the oscillation effects to correspond to the 9 respective reactor cores and complex at Taishan, Yangjiang and Daya Bay. Here each core $r$ produces a relative share $\phi_r$ of the total flux, determined by its fission rate and distance to the JUNO detector. The final forward model for the expected number of events per reconstructed energy bin is therefore
\begin{equation}
N_j = \int dE_\nu\, R_j(E_\nu)\,\sum_r \varphi_r(E_\nu)\,P_{ee}(E_\nu, L_r),
\end{equation}
%
The detector response $R$ maps each true neutrino energy to the reconstructed prompt-energy bins. It is built from the IBD kinematics, the positron energy non-linearity curve published by the JUNO collaboration, and a Gaussian energy resolution whose energy dependence is fitted to the published calibration-source resolutions~\cite{JUNO:2025first,JUNO:2026tap}. Furthermore, in order to best reproduce the JUNO released results, we use the observed non-oscillated flux that JUNO reports~\cite{JUNO:2025first} to tune our predicted non-oscillated flux using one ratio parameter per reconstructed prompt energy bin, similar to the practice done in previous external analysis of the JUNO results~\cite{Esteban:2026phq}. The resulting event distribution in reconstructed prompt energy for both oscillated and non-oscillated cases and their comparisons are shown in \cref{fig:junofig3}. The lower panel shows the ratios of data over JUNO best fit prediction as well as ratio of our work's best fit prediction to JUNO's best fit prediction; the two best fit predictions agree within 1.0\% on average with a 2.4\% r.m.s.\ over the 156 bins.

\textbf{\textit{Systematic uncertainties.}} To reproduce the JUNO results, due to the lack of the exact covariance matrix presented in~\cite{JUNO:2026tap}, we choose to use a pure Poisson likelihood with 34 profiled nuisance parameters. These parameters and their priors are listed in \cref{tab:juno-systematics}.

\begin{figure}
    \centering
    \includegraphics[width=0.9\linewidth]{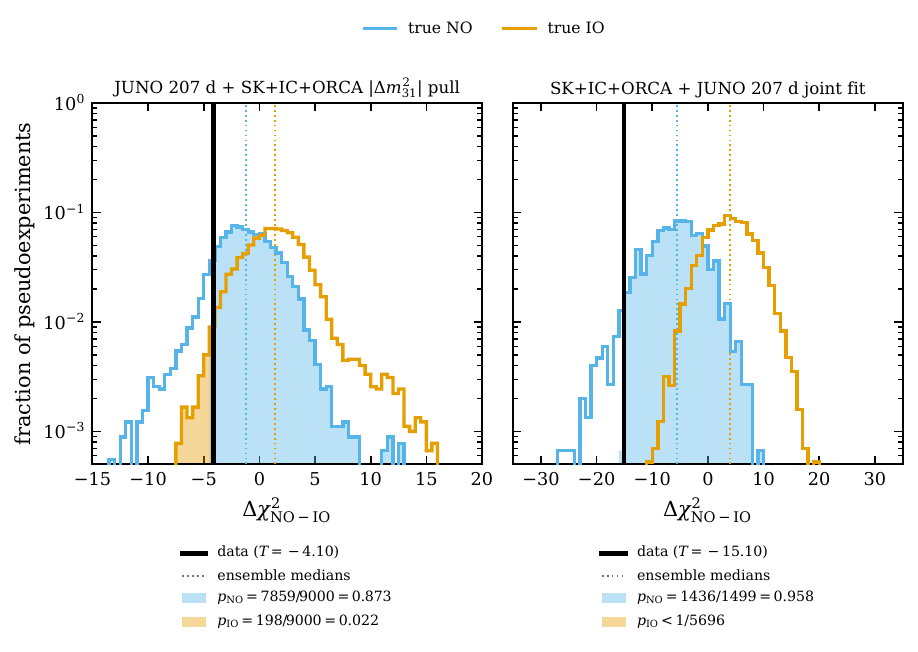}
    \caption{\textbf{Distributions of the ordering test statistic $\Delta\chi^2_{\mathrm{NO-IO}}$ for pseudoexperiments generated under the normal (blue) and inverted (orange) orderings}). \textbf{Left}: JUNO 207.2-day data alone, with the $|\Delta m^2_{31}|$ profile of the combined atmospheric fit entering as an external term. \textbf{Right}: the full joint fit of the atmospheric data and the JUNO 207.2-day data. Black line marks the observed $\Delta \chi^2_{\rm NO-IO}$ value and shaded areas mark the pseudoexperiments at least as extreme as the data; the resulting $p$-values are given in the legends.}
    \label{fig:juno-nmo}
\end{figure}

\begin{table}[t]
\centering
\caption{\textbf{\textit{Nuisance parameters of the JUNO fit.}} Every pull carries a unit Gaussian penalty and is
profiled at each oscillation point. Entries marked with an asterisk apply only to the TAO-flux
variant of the appendix, in which the two Taishan cores use the TAO-unfolded yield.
The geoneutrino entry is the
model prediction and its uncertainty, which the collaboration leaves free in its own fit. The
``other'' template collects accidentals, $^{13}$C($\alpha$,n)$^{16}$O, fast neutrons, double
neutrons and atmospheric neutrinos, whose published uncertainties sum in quadrature to $34\,\%$;
a $50\,\%$ width is assigned.}
\label{tab:juno-systematics}
\begin{tabular}{llcl}
\hline\hline
Class & Parameter & Pulls & Prior \\
\hline
Normalisation & overall rate & 1 & $2.4\,\%$ (59.1\,d), $2.8\,\%$ (207.2\,d) \\
\hline
Flux shape & Daya Bay modes & 25 & eigenmodes from covariance \\
           & TAO modes$^{*}$ & 9 & eigenmodes from covariance \\
\hline
Detector   & energy scale & 1 & $0.5\,\%$ \\
           & non-linearity & 1 & published band \\
           & resolution & 1 & $5\,\%$ \\
\hline
Backgrounds & $^{9}$Li/$^{8}$He & 1 & $26\,\%$ of $1.05$ (33\,\% of $4.3$) \\
            & geoneutrinos & 1 & $40\,\%$ of $1.30$ (42\,\% of $1.2$) \\
            & world reactors & 1 & $10\,\%$ of $0.92$ (10\,\% of $0.88$) \\
            & $^{214}$Bi--$^{214}$Po & 1 & $74\,\%$ of $0.42$ (56\,\% of $0.18$) \\
            & other & 1 & $50\,\%$ of $0.33$ \\
\hline
Total & & 34 (43$^{*}$) & \\
\hline\hline
\end{tabular}
\end{table}

\begin{figure}
    \centering
    \includegraphics[width=0.9\linewidth]{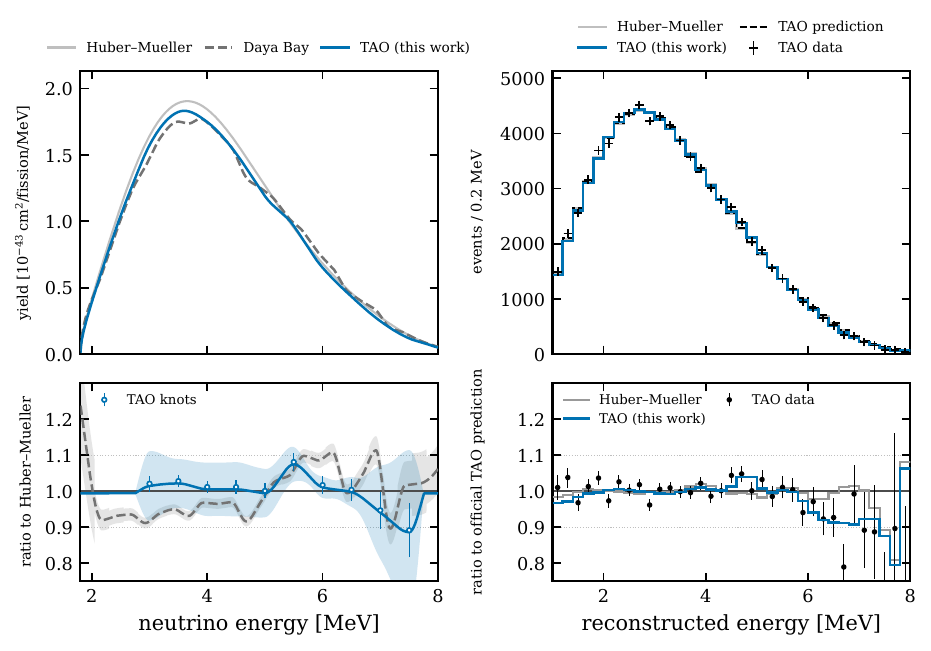}
    \caption{\textbf{\textit{Reactor antineutrino flux anchors.}} \textbf{Left}: IBD yield per fission as a function of true neutrino energy for the Huber–Mueller prediction, the Daya Bay unfolded measurement used for all cores in the production fit, and the TAO-anchored yield of this work used for the Taishan cores in the alternative fit, with the lower panel showing the shape ratio of the two measurements to Huber–Mueller, the one-sigma envelope of the corresponding shape pulls, and the ten fitted TAO knots with their statistical uncertainties. \textbf{Right}: the published TAO prompt spectrum together with Huber–Mueller, the TAO-anchored yield and the official TAO prediction, each folded through our model of the TAO detector response and normalised to the data over 2–7 MeV; the lower panel shows the ratio of the data and of the two folded yields to the official prediction.}
    \label{fig:junotao}
\end{figure}

\textbf{\textit{59-day-fit and TAO-flux-included-207-day-fit.}} We also consider two alternative implementations of the fit. The first is a simple analysis using the published 59 days data~\cite{JUNO:2025first}. The second one uses the recently published TAO flux measurements~\cite{2026junotao} for the Taishan reactor cores while keeping the Daya Bay flux measurement for the other cores. Since we tune the predicted non-oscillated unfolded flux to match the JUNO release, adding in the TAO measurements does not change the nominal reconstructed prompt energy flux at no oscillation; it instead changes the true neutrino energy shape as well as the nuisance parameters. Here we elaborate on how we implement the TAO measurements. To implement the TAO flux constraints, we unfold the non-oscillation spectrum in reconstructed prompt energy to the true neutrino energy by finding the corrections $s(E_\nu)$ to the Huber-Muller flux at 10 knots from 3.3 MeV to 7 MeV such that 

\begin{equation}
\varphi_{\rm TAO}(E_\nu) = \varphi_{\rm HM}(E_\nu)\,e^{s(E_\nu)},
\end{equation}

and fitting a cubic spline to it. We fit for these $s(E_\nu)$ such that when convolved with the response matrix in the forward model we reproduce the spectrum in reconstructed energy. The response matrix is built from positron non-linearity, energy resolution and leakage information all provided in~\cite{2026junotao}.
Our results are compared with the Huber–Mueller theoretical prediction in terms of the true neutrino energy and with the official reconstructed-energy distribution~\cite{2026junotao} in~\Cref{fig:junotao}.

\begin{figure}
    \centering
    \includegraphics[width=0.9\linewidth]{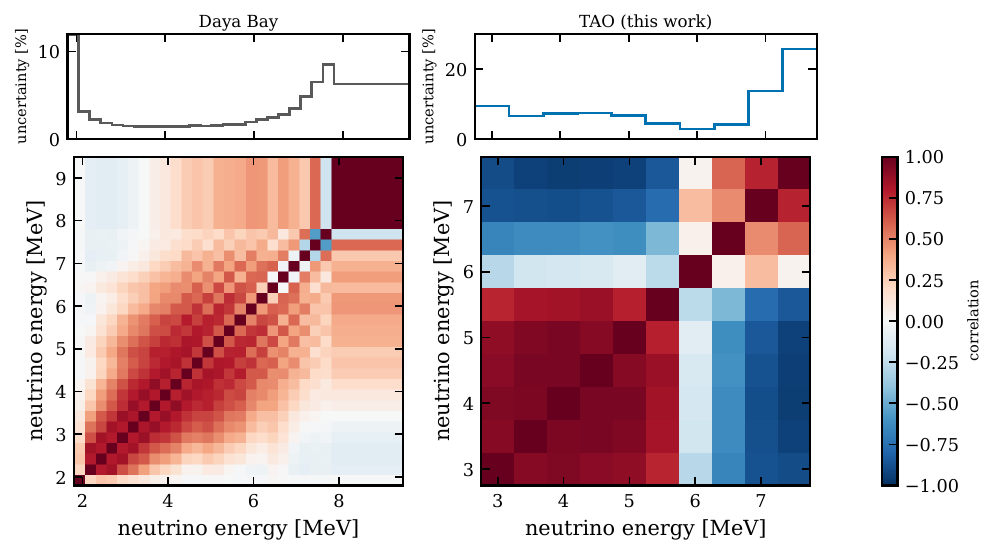}
    \caption{\textbf{\textit{Correlation matrices of the two reactor flux anchors in true neutrino energy.}} \textbf{Left}: the Daya Bay unfolded yield, from the covariance released with the measurement over its 25 energy bins. \textbf{Right}: the TAO-anchored correction of this work over its ten knots, with the covariance obtained from the unfolding fit, combining the statistical uncertainty of the published spectrum with the published systematic components propagated through the fit. The panels above each matrix show the relative uncertainty per bin; the eigenvectors of each matrix define the flux-shape nuisance parameters of the respective fit.}
    \label{fig:junocovariance}
\end{figure}

It is also appropriate to further discuss the nuisance parameter implementation here. For both cases of Daya Bay flux and TAO flux, we use the covariance matrices of the energy bins in unfolded true neutrino energy, and extract the eigenvalue and eigenvector information. Each nuisance parameter $\eta_k$ multiplies the flux by $1 + \eta_k\sqrt{\lambda_k},u_k(E_\nu)$, where $u_k$ is the $k$-th eigenvector of the covariance interpolated onto the energy grid and $\lambda_k$ its eigenvalue, so that unit Gaussian pulls on the $\eta_k$ reproduce the covariance exactly. We also present the covariance matrix of the official Daya Bay release~\cite{DayaBay:2025ngb} alongside our work's derived covariance for TAO in ~\cref{fig:junocovariance}.

\begin{figure}
    \centering
    \includegraphics[width=0.9\linewidth]{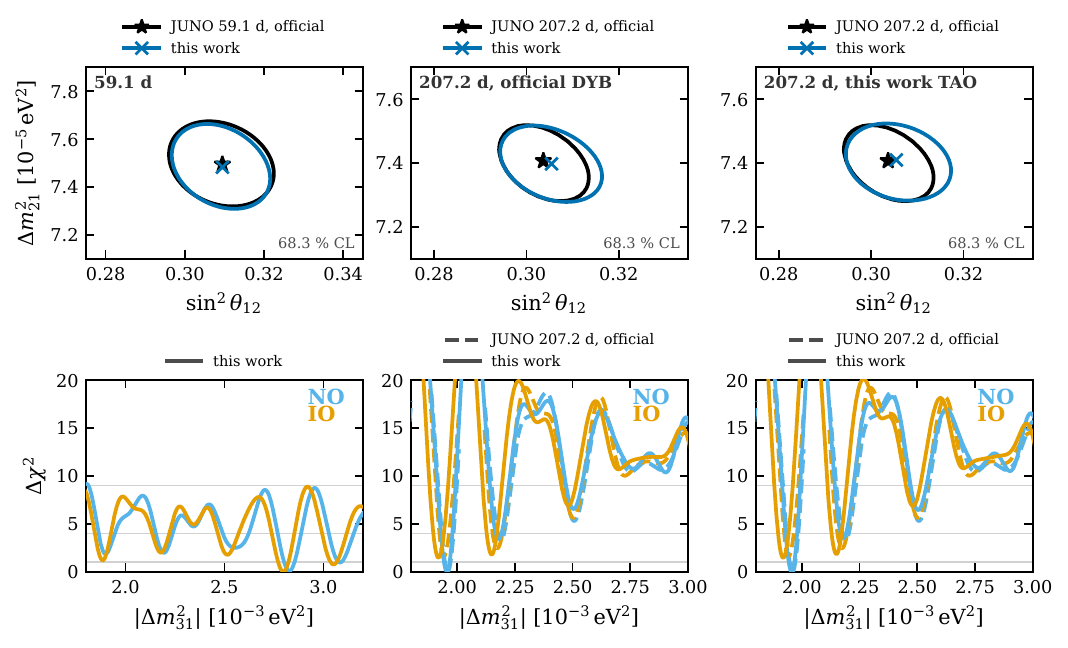}
    \caption{\textbf{\textit{Reproduction of the JUNO reactor results with our forward model.}} Each column compares one configuration with the official result only: the 59.1-day data set (\textbf{left}), the 207.2-day data set with the Daya Bay flux for all cores (\textbf{middle}), and the 207.2-day data set with the TAO flux for the two Taishan cores (\textbf{right}). \textbf{Top row}: 68.3 \% CL regions in the solar plane, with the official best fit (star) and ours (cross); \textbf{Bottom row}: $\Delta \chi^2$ as a function of $|\Delta m^2_{31}|$ for the normal and inverted orderings, each referenced to its own global minimum.}
    \label{fig:junosolardm2}
\end{figure}

\textbf{\textit{Fit results and Validation.}} Finally, we present the solar oscillation parameters and the $\Delta m^2_{31}$ fit results from all three setups in \cref{fig:junosolardm2}. Here we can see that the 59.1-day arm reproduces the released solar-parameter best fit almost completely, which is expected as we take the data release files directly. For the 207-day results reproduction, we see that both the Daya-Bay-only flux modeling and the TAO-included flux modeling reach good agreement with the official contour from \cite{JUNO:2026tap}, where the contour is larger for the TAO-included arm, an expected outcome due to the flux systematics having more systematics freedom. We therefore choose to report the study with the Daya-Bay-only flux modeling. If we were to use the TAO-included version, we observe an increase of $\Delta \chi^2_{\rm NO-IO} = -15.6$, but close enough to the results quoted in the main text. 

Using this simulation, we have obtained the NMO test statistics whose pseudoexperiment distribution is shown in \cref{fig:juno-nmo}. In the case of JUNO fit with our combined atmospheric $\Delta m^2_{31}$ pull term we observe a 2$\sigma$ preference for NO; in a full combined fit, we expect an evidence for NO at $3.9 \sigma$, where our pseudoexperiments do not find a tail event yet, but show a clear gaussian distribution to support the $3.9 \sigma$ claim.

\clearpage

\clearpage   
{\footnotesize\renewcommand{\arraystretch}{1.02}\setlength{\tabcolsep}{5pt}
\begin{longtable}{@{}l l r r r r@{}}
\caption{\textbf{\textit{Nuisance parameters of the combined fit.}}}\label{tab:systs}\label{table:FluxSysts}\label{table:XSecSysts}\\
\toprule
Parameter & Experiments & Central value & Prior $1\sigma$ & Best fit (NO) & Pull [$\sigma$] \\
\midrule\endfirsthead
\toprule
Parameter & Experiments & Central value & Prior $1\sigma$ & Best fit (NO) & Pull [$\sigma$] \\
\midrule\endhead
\midrule\multicolumn{6}{r}{\emph{continued on the next page}}\\\endfoot
\bottomrule\endlastfoot
\multicolumn{6}{@{}l}{\textbf{Atmospheric neutrino flux}}\\*
Normalization, $E_\nu<1$\,GeV & SK, IC, ORCA & 1.00 & 0.25 & 1.05 & $+0.18$ \\
Normalization, $E_\nu>1$\,GeV & SK, IC, ORCA & 1.00 & 0.15 & 0.97 & $-0.20$ \\
Spectral index & SK, IC, ORCA & 0.00 & 0.1 & -0.05 & $-0.45$ \\
Up-going zenith shape & SK, IC, ORCA & 0.00 & 0.01 & 0.00 & $-0.05$ \\
Down-going zenith shape & SK, IC, ORCA & 0.00 & 0.01 & 0.00 & $+0.04$ \\
Solar modulation & SK, IC, ORCA & 0.00 & 0.15 & -0.08 & $-0.56$ \\
\rowcolor[gray]{0.91}$K/\pi$ production ratio & SK, IC, ORCA & 0.00 & 0.1 & 0.14 & $+1.43$ \\
Horizontal/vertical ratio & SK, IC, ORCA & 0.00 & 0.03 & -0.002 & $-0.07$ \\
$\bar\nu_e/\nu_e$ ratio, $E_\nu<1$\,GeV & SK, IC, ORCA & 1.00 & 0.05 & 0.975 & $-0.49$ \\
$\bar\nu_e/\nu_e$ ratio, $1<E_\nu<10$\,GeV & SK, IC, ORCA & 1.00 & 0.05 & 0.982 & $-0.36$ \\
$\bar\nu_e/\nu_e$ ratio, $E_\nu>10$\,GeV & SK, IC, ORCA & 1.00 & 0.08 & 1.030 & $+0.38$ \\
$\nu_\mu/\nu_e$ ratio, $E_\nu<1$\,GeV & SK, IC, ORCA & 1.00 & 0.02 & 1.00 & $-0.01$ \\
$\nu_\mu/\nu_e$ ratio, $1<E_\nu<10$\,GeV & SK, IC, ORCA & 1.00 & 0.03 & 0.989 & $-0.38$ \\
$\nu_\mu/\nu_e$ ratio, $E_\nu>10$\,GeV & SK, IC, ORCA & 1.00 & 0.05 & 1.040 & $+0.80$ \\
$\bar\nu_\mu/\nu_\mu$ ratio, $E_\nu<1$\,GeV & SK, IC, ORCA & 1.00 & 0.02 & 1.001 & $+0.03$ \\
$\bar\nu_\mu/\nu_\mu$ ratio, $1<E_\nu<10$\,GeV & SK, IC, ORCA & 1.00 & 0.06 & 0.953 & $-0.79$ \\
$\bar\nu_\mu/\nu_\mu$ ratio, $E_\nu>10$\,GeV & SK, IC, ORCA & 1.00 & 0.15 & 1.03 & $+0.20$ \\
\midrule\multicolumn{6}{@{}l}{\textbf{Neutrino interaction cross sections}}\\*
$\nu_\tau$ CC & SK, IC, ORCA & 1.00 & 0.25 & 0.77 & $-0.94$ \\
NC/CC ratio & SK, IC, ORCA & 1.00 & 0.2 & 1.01 & $+0.05$ \\
NC hadron production & SK, IC, ORCA & 1.00 & 0.1 & 1.00 & $+0.02$ \\
Axial mass $M_A$ & SK, IC, ORCA & 1.00 & 0.1 & 1.01 & $+0.08$ \\
\rowcolor[gray]{0.91}DIS & SK, IC, ORCA & 1.00 & 0.05 & 1.072 & $+1.44$ \\
\rowcolor[gray]{0.78}CCQE normalization & SK, IC, ORCA & 1.00 & 0.1 & 0.75 & $-2.48$ \\
CCQE $\bar\nu/\nu$ & SK, IC, ORCA & 1.00 & 0.1 & 0.90 & $-0.96$ \\
CCQE $\nu_\mu/\nu_e$ & SK, IC, ORCA & 1.00 & 0.1 & 1.01 & $+0.15$ \\
CC1$\pi$ production & SK, IC, ORCA & 1.00 & 0.1 & 0.91 & $-0.93$ \\
CC1$\pi$ $\bar\nu_e/\nu_e$ & SK, IC, ORCA & 1.00 & 0.1 & 0.97 & $-0.29$ \\
\rowcolor[gray]{0.91}CC1$\pi$ $\bar\nu_\mu/\nu_\mu$ & SK, IC, ORCA & 1.00 & 0.1 & 0.84 & $-1.59$ \\
CCQE energy dependence, $E_\nu<1.3$\,GeV & SK, IC, ORCA & 0.00 & 0.4 & 0.36 & $+0.91$ \\
\rowcolor[gray]{0.91}CCQE $\nu_e$, $E_\nu>1$\,GeV & SK, IC, ORCA & 1.00 & 0.25 & 0.74 & $-1.04$ \\
\rowcolor[gray]{0.91}CCQE $\nu_\mu$, $E_\nu>1$\,GeV & SK, IC, ORCA & 1.00 & 0.25 & 0.69 & $-1.23$ \\
CCQE energy dependence, $E_\nu>1.3$\,GeV & SK, IC, ORCA & 0.00 & 0.2 & -0.17 & $-0.87$ \\
2p2h normalization & SK, IC, ORCA & 1.00 & 0.1 & 0.98 & $-0.21$ \\
2p2h $\bar\nu/\nu$ & SK, IC, ORCA & 1.00 & 0.1 & 1.00 & $-0.02$ \\
2p2h $\nu_\mu/\nu_e$ & SK, IC, ORCA & 1.00 & 0.1 & 1.04 & $+0.42$ \\
CC1$\pi$ $\pi^0/\pi^\pm$ & SK, IC, ORCA & 1.00 & 0.4 & 1.18 & $+0.45$ \\
Coherent $\pi$ production & SK, IC, ORCA & 1.00 & 1 & 1.4 & $+0.41$ \\
1p1h $\nu_e$, $E_\nu<1$\,GeV & SK, IC, ORCA & 1.00 & 0.05 & 0.972 & $-0.55$ \\
\midrule\multicolumn{6}{@{}l}{\textbf{Super-Kamiokande detector}}\\*
PC reduction, SK-I & SK & 1.00 & 0.024 & 0.984 & $-0.66$ \\
PC reduction, SK-II & SK & 1.00 & 0.048 & 0.965 & $-0.73$ \\
PC reduction, SK-III & SK & 1.00 & 0.009 & 0.9992 & $-0.08$ \\
PC reduction, SK-IV--V & SK & 1.00 & 0.01 & 0.993 & $-0.70$ \\
Multi-GeV non-$\nu$ background, SK-I & SK & 1.00 & 0.13 & 0.96 & $-0.30$ \\
Multi-GeV non-$\nu$ background, SK-II & SK & 1.00 & 0.38 & 0.94 & $-0.17$ \\
Multi-GeV non-$\nu$ background, SK-III & SK & 1.00 & 0.27 & 0.91 & $-0.33$ \\
Multi-GeV non-$\nu$ background, SK-IV--V & SK & 1.00 & 0.18 & 0.96 & $-0.20$ \\
FC reduction, SK-I & SK & 1.00 & 0.002 & 0.9999 & $-0.03$ \\
FC reduction, SK-II & SK & 1.00 & 0.002 & 1.00 & $-0.02$ \\
FC reduction, SK-III & SK & 1.00 & 0.008 & 0.9996 & $-0.05$ \\
FC reduction, SK-IV--V & SK & 1.00 & 0.013 & 1.00 & $0.00$ \\
Sub-GeV 2-ring $\pi^0$ selection, SK-I & SK & 1.00 & 0.056 & 0.990 & $-0.17$ \\
Sub-GeV 2-ring $\pi^0$ selection, SK-II & SK & 1.00 & 0.044 & 0.997 & $-0.07$ \\
Sub-GeV 2-ring $\pi^0$ selection, SK-III & SK & 1.00 & 0.059 & 0.996 & $-0.07$ \\
Sub-GeV 2-ring $\pi^0$ selection, SK-IV--V & SK & 1.00 & 0.056 & 0.975 & $-0.45$ \\
Sub-GeV 1-ring $\pi^0$ selection, SK-I & SK & 1.00 & 0.12 & 0.99 & $-0.05$ \\
Sub-GeV 1-ring $\pi^0$ selection, SK-II & SK & 1.00 & 0.12 & 1.00 & $-0.04$ \\
Sub-GeV 1-ring $\pi^0$ selection, SK-III & SK & 1.00 & 0.1 & 1.00 & $-0.01$ \\
Sub-GeV 1-ring $\pi^0$ selection, SK-IV--V & SK & 1.00 & 0.15 & 0.98 & $-0.13$ \\
Multi-ring $\nu/\bar\nu$ separation, SK-I & SK & 1.00 & 0.072 & 1.037 & $+0.51$ \\
Multi-ring $\nu/\bar\nu$ separation, SK-II & SK & 1.00 & 0.079 & 1.027 & $+0.34$ \\
Multi-ring $\nu/\bar\nu$ separation, SK-III & SK & 1.00 & 0.077 & 1.012 & $+0.16$ \\
\rowcolor[gray]{0.91}Multi-ring $\nu/\bar\nu$ separation, SK-IV--V & SK & 1.00 & 0.068 & 1.080 & $+1.17$ \\
Multi-ring $e/\mu$ separation, SK-I & SK & 1.00 & 0.06 & 0.998 & $-0.04$ \\
Multi-ring $e/\mu$ separation, SK-II & SK & 1.00 & 0.038 & 0.999 & $-0.02$ \\
Multi-ring $e/\mu$ separation, SK-III & SK & 1.00 & 0.053 & 1.00 & $0.00$ \\
Multi-ring $e/\mu$ separation, SK-IV--V & SK & 1.00 & 0.033 & 0.999 & $-0.04$ \\
Multi-ring $e$/other separation, SK-I & SK & 1.00 & 0.057 & 0.986 & $-0.25$ \\
Multi-ring $e$/other separation, SK-II & SK & 1.00 & 0.041 & 0.993 & $-0.16$ \\
Multi-ring $e$/other separation, SK-III & SK & 1.00 & 0.049 & 0.996 & $-0.09$ \\
Multi-ring $e$/other separation, SK-IV--V & SK & 1.00 & 0.034 & 0.979 & $-0.63$ \\
PC stopping/through-going separation, SK-I & SK & 1.00 & 0.25 & 1.10 & $+0.39$ \\
PC stopping/through-going separation, SK-II & SK & 1.00 & 0.14 & 1.01 & $+0.08$ \\
PC stopping/through-going separation, SK-III & SK & 1.00 & 0.3 & 1.05 & $+0.18$ \\
PC stopping/through-going separation, SK-IV--V & SK & 1.00 & 0.1 & 1.04 & $+0.41$ \\
$\pi^0$ ring separation, SK-I & SK & 1.00 & 0.023 & 1.00 & $+0.01$ \\
$\pi^0$ ring separation, SK-II & SK & 1.00 & 0.023 & 1.00 & $0.00$ \\
$\pi^0$ ring separation, SK-III & SK & 1.00 & 0.023 & 1.00 & $+0.01$ \\
$\pi^0$ ring separation, SK-IV--V & SK & 1.00 & 0.03 & 1.002 & $+0.06$ \\
$e$-like ring separation, SK-I & SK & 1.00 & 0.035 & 0.999 & $-0.03$ \\
$e$-like ring separation, SK-II & SK & 1.00 & 0.038 & 1.00 & $0.00$ \\
$e$-like ring separation, SK-III & SK & 1.00 & 0.013 & 1.00 & $+0.01$ \\
$e$-like ring separation, SK-IV--V & SK & 1.00 & 0.018 & 0.998 & $-0.13$ \\
$\mu$-like ring separation, SK-I & SK & 1.00 & 0.045 & 0.994 & $-0.13$ \\
$\mu$-like ring separation, SK-II & SK & 1.00 & 0.08 & 0.978 & $-0.27$ \\
$\mu$-like ring separation, SK-III & SK & 1.00 & 0.026 & 1.001 & $+0.02$ \\
$\mu$-like ring separation, SK-IV--V & SK & 1.00 & 0.023 & 1.002 & $+0.09$ \\
Single-ring particle identification, SK-I & SK & 1.00 & 0.002 & 1.00 & $0.00$ \\
Single-ring particle identification, SK-II & SK & 1.00 & 0.007 & 1.0001 & $+0.01$ \\
Single-ring particle identification, SK-III & SK & 1.00 & 0.0026 & 1.00 & $0.00$ \\
Single-ring particle identification, SK-IV--V & SK & 1.00 & 0.0035 & 0.9996 & $-0.12$ \\
Multi-ring particle identification, SK-I & SK & 1.00 & 0.065 & 0.997 & $-0.05$ \\
Multi-ring particle identification, SK-II & SK & 1.00 & 0.1 & 0.99 & $-0.06$ \\
Multi-ring particle identification, SK-III & SK & 1.00 & 0.05 & 1.00 & $0.00$ \\
Multi-ring particle identification, SK-IV--V & SK & 1.00 & 0.035 & 0.999 & $-0.04$ \\
Up-$\mu$ showering separation, SK-I & SK & 1.00 & 0.0102 & 0.996 & $-0.37$ \\
Up-$\mu$ showering separation, SK-II & SK & 1.00 & 0.0132 & 0.997 & $-0.25$ \\
Up-$\mu$ showering separation, SK-III & SK & 1.00 & 0.0072 & 0.9993 & $-0.09$ \\
Up-$\mu$ showering separation, SK-IV--V & SK & 1.00 & 0.009 & 0.9927 & $-0.82$ \\
Up-$\mu$ stopping background, SK-I & SK & 1.00 & 0.007 & 1.0003 & $+0.05$ \\
Up-$\mu$ stopping background, SK-II & SK & 1.00 & 0.007 & 1.0002 & $+0.02$ \\
Up-$\mu$ stopping background, SK-III & SK & 1.00 & 0.007 & 1.0001 & $+0.02$ \\
Up-$\mu$ stopping background, SK-IV--V & SK & 1.00 & 0.005 & 1.0004 & $+0.08$ \\
Up-$\mu$ showering background, SK-I & SK & 1.00 & 0.005 & 0.9993 & $-0.13$ \\
Up-$\mu$ showering background, SK-II & SK & 1.00 & 0.005 & 0.9996 & $-0.07$ \\
Up-$\mu$ showering background, SK-III & SK & 1.00 & 0.005 & 0.9998 & $-0.05$ \\
Up-$\mu$ showering background, SK-IV--V & SK & 1.00 & 0.003 & 0.9994 & $-0.20$ \\
Up-$\mu$ non-showering background, SK-I & SK & 1.00 & 0.005 & 1.0012 & $+0.24$ \\
Up-$\mu$ non-showering background, SK-II & SK & 1.00 & 0.005 & 1.0006 & $+0.13$ \\
Up-$\mu$ non-showering background, SK-III & SK & 1.00 & 0.005 & 1.0004 & $+0.08$ \\
Up-$\mu$ non-showering background, SK-IV--V & SK & 1.00 & 0.003 & 1.0011 & $+0.36$ \\
\rowcolor[gray]{0.91}Fiducial volume & SK & 1.00 & 0.02 & 1.027 & $+1.36$ \\
Neutron tagging, sub-GeV & SK & 1.00 & 0.12 & 1.04 & $+0.31$ \\
Neutron tagging, multi-GeV & SK & 1.00 & 0.12 & 1.02 & $+0.19$ \\
Energy scale, SK-I & SK & 1.00 & 0.033 & 0.986 & $-0.44$ \\
Energy scale, SK-II & SK & 1.00 & 0.028 & 0.996 & $-0.15$ \\
Energy scale, SK-III & SK & 1.00 & 0.024 & 0.999 & $-0.05$ \\
Energy scale, SK-IV--V & SK & 1.00 & 0.021 & 1.004 & $+0.21$ \\
FC multi-GeV relative normalization & SK & 1.00 & 0.05 & 0.976 & $-0.49$ \\
Up-$\mu$ stopping background, horizontal, low $p$ & SK & 1.00 & 0.09 & 1.019 & $+0.21$ \\
Up-$\mu$ stopping background, horizontal, high $p$ & SK & 1.00 & 0.036 & 0.980 & $-0.55$ \\
Up-$\mu$ non-showering background, horizontal & SK & 1.00 & 0.013 & 1.009 & $+0.69$ \\
Up/down energy-scale asymmetry, SK-I & SK & 0.00 & 0.006 & 0.0013 & $+0.21$ \\
Up/down energy-scale asymmetry, SK-II & SK & 0.00 & 0.006 & 0.0007 & $+0.12$ \\
Up/down energy-scale asymmetry, SK-III & SK & 0.00 & 0.013 & 0.002 & $+0.16$ \\
Up/down energy-scale asymmetry, SK-IV--V & SK & 0.00 & 0.005 & 0.0039 & $+0.78$ \\
$\bar\nu_e$-like neutron-tag migration, sub-GeV & SK & 1.00 & 0.1 & 1.08 & $+0.76$ \\
$\bar\nu_e$-like neutron-tag migration, multi-GeV & SK & 1.00 & 0.1 & 1.09 & $+0.92$ \\
Decay-electron tagging, SK-I & SK & 1.00 & 0.1 & 0.98 & $-0.19$ \\
Decay-electron tagging, SK-II & SK & 1.00 & 0.1 & 1.00 & $-0.04$ \\
Decay-electron tagging, SK-III & SK & 1.00 & 0.1 & 0.99 & $-0.10$ \\
Decay-electron tagging, SK-IV--V & SK & 1.00 & 0.1 & 1.05 & $+0.48$ \\
$e$-like ring counting, low $p$, SK-I--III & SK & 1.00 & 0.04 & 1.024 & $+0.59$ \\
FC/PC separation & SK & 1.00 & 0.01 & 1.002 & $+0.19$ \\
\midrule\multicolumn{6}{@{}l}{\textbf{IceCube DeepCore detector}}\\*
DOM efficiency & IC & 1.00 & 0.1 & 1.03 & $+0.35$ \\
Hole ice $p_0$ & IC & 0.10 & 0.1 & 0.01 & $-0.90$ \\
Hole ice $p_1$ & IC & -0.05 & 0.1 & -0.10 & $-0.51$ \\
Bulk-ice absorption & IC & 1.00 & 0.1 & 0.94 & $-0.64$ \\
Bulk-ice scattering & IC & 1.00 & 0.1 & 1.03 & $+0.28$ \\
\midrule\multicolumn{6}{@{}l}{\textbf{KM3NeT/ORCA6 detector}}\\*
Overall normalization & ORCA & 1.00 & 0.2 & 0.97 & $-0.15$ \\
High-purity-track normalization & ORCA & 1.00 & 0.2 & 0.95 & $-0.27$ \\
Shower normalization & ORCA & 1.00 & 0.2 & 0.98 & $-0.12$ \\
$\nu_\tau$ CC event selection normalization & ORCA & 1.00 & 0.2 & 0.99 & $-0.03$ \\
High-energy normalization & ORCA & 1.00 & 0.5 & 1.28 & $+0.56$ \\
\end{longtable}}

\end{document}